\documentclass[12pt]{iopart}
\usepackage[backend=biber,sorting=none]{biblatex}
\expandafter\let\csname equation*\endcsname\relax
\expandafter\let\csname endequation*\endcsname\relax
\usepackage{mathtools}

\usepackage[toc,page]{appendix}
\usepackage{graphicx}
\usepackage{listings}
\usepackage[T1]{fontenc}
\usepackage[utf8]{inputenc}
\usepackage[nottoc,numbib]{tocbibind}
\usepackage{setspace}
\usepackage{float}
\usepackage[dvipsnames]{xcolor}
\graphicspath{{images/}}

\begin{document}
\title[Controlling enhanced ionisation with machine learning]{Controlling quantum effects in enhanced strong-field ionisation with machine-learning techniques}
\author{H Chomet, S Plesnik, C Nicolae, J Dunham, L Gover, T Weaving and C Figueira de Morisson Faria}
\address{Department of Physics and Astronomy, University College London, Gower Street, London WC1E 6BT, UK}
\date{\today}
\begin{abstract}
We study non-classical pathways and quantum interference in enhanced ionisation of diatomic molecules in strong laser fields using machine learning techniques. Quantum interference provides a ‘bridge’, which facilitates intramolecular population transfer. Its frequency is higher than that of the field, intrinsic to the system and depends on several factors, for instance the state of the initial wavepacket or the internuclear separation. Using dimensionality reduction techniques, namely t-distributed stochastic neighbour embedding (t-SNE) and principal component analysis (PCA),  we investigate the effect of multiple parameters at once and find optimal conditions for enhanced ionisation in static fields, and  controlled ionisation release for two-colour driving fields. This controlled ionisation manifests itself as a step-like behaviour in the time-dependent autocorrelation function. We explain the features encountered with phase-space arguments, and also establish a hierarchy of parameters for controlling ionisation via phase-space Wigner quasiprobability flows, such as specific coherent superpositions of states, electron localisation and internuclear-distance ranges.  
\end{abstract}
%\keywords{attosecond science, machine learning, strong field ionisation, enhanced ionisation, phase space}

\maketitle

\section{Introduction}

Manipulating coherent superpositions of quantum states and nonclassical pathways has been a central question to many areas of science, such as coherent control \cite{Shapiro2000,Bauerle2018,Hikosaka2019} and quantum information \cite{Reiserer2015}. Important applications include controlling chemical reactions \cite{Peller2020} or electron dynamics in ultrafast molecular dissociation \cite{Kling2006,Sansone2010,Singh2010,Xu2017}, creating quantum switches \cite{Cirac1997,Kimble2008,Reiserer2015} and enhancing nonclassicality in extended systems such as light-harvesting compounds \cite{Wilde2009,O’Reilly2014}. In this wide range of scenarios, decoherence must be kept at bay, so that the timescales of interest are much shorter than the decoherence times. This brings one's attention to how coherent superpositions of quantum states and nonclassical pathways may be controlled in attoscience. 

Attoscience has emerged from the interaction of matter with very intense laser fields, typically of the order of $10^{13}\mathrm{W}/\mathrm{cm}^2$ and deals with some of the shortest time scales in nature, of the order of $10^{-18}s$. These extremely short timescales bring about the possibility of controlling real-time electron dynamics \cite{Krausz2009,Lepine2014}, and have triggered many applications. Examples are subfemtosecond imaging of matter (for reviews see, e.g., \cite{Lein2007,Salieres2012R}), high-harmonic spectroscopy \cite{Marangos2016}, ultrafast photoelectron holography \cite{Huismans2011,Faria2020}, attosecond electron or hole migration \cite{Smirnova2009,Mairesse2010,Calegari2014,Kuleff2016,Calegari2016Migration}, and, more recently, strong-field phenomena in solids \cite{Ghimire2014,Vampa2018,Yu2019,Ortmann2021} and nanostructures \cite{Ciappina2017,Ciappina2019}. It has even been speculated that, in the future, laser-induced electron dynamics may lead to optoelectronic computers with switching rates 100,000 times higher than existing digital electronic systems \cite{Schiffrin2013,Schultze2013,sommer2016attosecond}.

Recently, it has been shown that quantum interference and nonclassical pathways play a vital role in strong-field enhanced ionisation of diatomic molecules \cite{chomet2019quantum}. Enhanced ionisation is known for over three decades \cite{zuo1995charge,Seideman1995}, and consists in an increase of at least one order of magnitude in the ionisation rate of a stretched molecule, in comparison with that of an atom with a similar ionisation potential. It has been attributed to the narrowing of the upfield potential barrier due to the presence of an adjacent, downfield potential well, and to coupled charge-enhanced resonant states. Although the phenomenon is widely known, phase-space tools, such as the Wigner quasi-probability distributions \cite{wigner1932on} have shed new light on its behaviour \cite{he2008strong,takemoto2011time,chomet2019quantum}. Phase-space quasi-probability distributions have found enormous success in many research areas such as quantum optics \cite{schleich2011quantum,Barnett2005,leonhardt2010essential}, quantum information \cite{Braunstein2005,Serafini2017}, chemical physics \cite{Miller2001,Miller2005} and cold gases \cite{Blakie2008}, but are hitherto underused in attoscience (for a recent review on the overall landscape within this research field see our article \cite{chomet2021attoscience}).  Nevertheless, they provide valuable insight into the dynamics of these systems of interest and have been used to study phenomena such as strong-field ionisation \cite{czirjak2000wigner,he2008strong,takemoto2011time,zagoya2014quantum}, rescattering \cite{Graefe2012,Baumann2015} and entanglement \cite{Czirjak2013}. For recent work investigating different pathways in enhanced ionisation see \cite{Xu2015,Liu2021a}.

In 2008, \textit{momentum gates} occurring in phase space under strong oscillating laser fields were identified \cite{he2008strong}. These gates facilitate the flow of quasi-probability from one molecular centre to the other causing many ionisation bursts within a field cycle. These bursts were later attributed to a non-adiabatic response to the time-dependent field gradients \cite{Takemoto2010,takemoto2011time}. However, our previous publication \cite{chomet2019quantum} dispelled this notion, by showing that these momentum gates are present when there is no time dependence on the field \cite{chomet2019quantum}, or even no field at all \cite{kufel2020alternative}, and are primarily caused by quantum interference. The term \textit{quantum bridge} was coined to describe this mechanism. 

In the present article, we aim at controlling the pathways behind enhanced ionisation, which have been identified in \cite{chomet2019quantum}. Some pathways may be understood within a quasi-static, semiclassical picture, whereby the quasiprobability flows along the field gradient, with tails following equienergy curves \cite{balazs1990wigner, czirjak2000wigner, zagoya2014quantum}. Other pathways are enabled by quantum interference, which creates a bridge and provides a passage for direct intra-molecular quasi-probability flow. For a time-dependent field, one must bear in mind that the phase-space configuration also changes as time progresses. Therefore, the field can act both in favour of the cyclic motion of the quantum bridges (and the enhanced ionisation) or against it. These different temporal behaviours depend on a wealth of parameters and may be explored in order to control strong-field ionisation. Recently, it has been shown that quantum interference plays an important role even in the semiclassical ionisation pathway \cite{hack2021quantum_wigner}. 

Previous analyses focused on understanding the physical mechanisms behind the quantum bridges, and were restricted to a homonuclear molecular potential in the field-free setting, or under a strong static or monochromatic laser field. The initial wavepacket was taken to be an upfield or downfield localised Gaussian, or a symmetric coherent superposition thereof known as a cat state \cite{schleich2011quantum}. Here, we focus on how the quantum bridges can be manipulated by using  different relative phases and wavepacket localisations in the initial superposition state, molecular potentials of heteronuclear type, where each well is weighted asymmetrically, and electric field configurations. 
This is a complex task: With a wealth of tunable parameters, such as internuclear separation, relative phase and wavepacket localisation in the initial wave packet, differing nuclear charges, laser intensity and frequency as well as the pulse shape, it necessitates optimisation techniques. Machine learning methods have been explored in a wide range of scenarios in quantum physics (see the reviews \cite{Dunjko2018,Carleo2019,Carrasquilla2020} and the recent tutorial \cite{Dawid2022}) and ultrafast photonics \cite{Genty2021}. Recently, they have been employed in the attosecond science context to control attosecond pulses \cite{BelmiroChu2021}, retrieve the structure of a large molecule from laser-induced electron diffraction patterns \cite{Liu2021b} and to track quantum pathways within a spatio-temporal Feynman path integral framework \cite{Liu2020}.

In this work, we employ machine-learning techniques for dimensionality reduction, such as the T-Distributed Stochastic Neighbour Embedding (t-SNE) \cite{vandermaaten08a:tsne_original_paper} and Principal-Component Analysis (PCA) \cite{Pearson1901} to find overall trends and patterns that allow us to establish a hierarchy of parameters and classify different regimes for strong-field ionisation. Throughout, we are aided by a qualitative phase-space analysis, which allows us to understand the physics behind the patterns encountered.  We first employ a model molecule in a static field as a proof of concept, and, subsequently, use time-dependent two-colour driving fields. The autocorrelation function exhibit several distinct behaviours in time, including a stepwise profile in which ionisation is switched on or off. We also provide a physical interpretation for these features. 

This article is organised as follows. In section \ref{sec:model} we discuss our model, including its general features (section \ref{sec:features}) and  the physical quantities of interest (section \ref{sec:tools}). Section \ref{sec:dimReduction} is devoted to the dimensionality reduction methods and their implementation. Subsequently, in section \ref{sec:static} we provide a qualitative analysis (\ref{sec:qualitative}) and a proof of concept (\ref{sec:staticTSNE}) for static fields, and in section\ref{sec:TD} we show how strong-field ionisation in two-colour fields can be optimised to produce a stepwise temporal behaviour. We commence by examining the different clusters that result from the t-SNE (section \ref{sec:TDtsneResults}) and analyse them using physical arguments (section \ref{sec:TDfullAnalysis}). Finally, in section \ref{sec:conclusions} we summarize our results and provide possible outlooks for the present studies. 
%%%%%%%%%%%%%%%%%%%%%%%%%%%%%%%%%%%%%%%%%%%%%%%%%%%%%%%%%%%%%%%%%%%%%%%%%%%%%%

\section{Model}
\label{sec:model}
\subsection{General features}
\label{sec:features}
Our system is a one-dimensional single-electron molecule whose wave function dynamics $\Psi(x,t)$ are obtained by solving the time-dependent Schr\"odinger equation (TDSE) in atomic units using the split operator method \cite{feit1982solution}
\begin{equation}\label{eq:tdse}
    i \frac{\partial}{\partial t} \Psi(x,t) = 
    \left(
    - \frac{1}{2} \frac{\partial^2}{\partial x^2} + V(x) + x E(t)
    \right)
    \Psi(x, t).
\end{equation}
The quantity $E(t)$ is the external laser field and $V(x)$ is the molecular binding potential given by
\begin{equation}
\label{eq:potential}
    V(x) = Z_{\mathrm{r}} V_0(x-R/2) + Z_{\mathrm{l}} V_0(x+R/2),
\end{equation}
where $R$ is the internuclear distance. The first and second term appearing on the right-hand side of equation (\ref{eq:potential}) shall be called the left and right potential wells, respectively. Each potential well $V_0(x)$ is chosen to have soft-core form 
\begin{equation}\label{eq:softcore}
    V_0(x) = -\frac{1}{\sqrt{x^2 + a}},
\end{equation}
where $a$ is known as the softening parameter. This removes the singularity at $x=0$ of the true Coulomb potential, yet 
remains long-range \cite{javanainen1988numerical,eberly1990scale, su1991model}. This parameter is typically chosen such that the ionisation energy coincides with the ground state of the potential \cite{rae1994saturation}; throughout the following analysis $a=1$ is chosen. Using a Gaussian wavepacket (Eq.~(\ref{eq:gaussian_i})) with $p_0 = q_0 = 0.0$ and $\gamma = 0.5$, this ground state energy is ${{E}_{{\rm sc}}}=-0.67$ a.u..

The homonuclear case is recovered by setting $Z_{\mathrm{l}}=Z_{\mathrm{r}}=1$. The symbol $Z$ is chosen in reference to nuclear charge; however, in this toy model, $Z_i$ is allowed to be a continuous variable. Eq.~(\ref{eq:softcore}) provides qualitative insight into the behaviour of Coulomb potentials, which suffices for the objectives of the present work. Recently, however, the soft-core potential has been modified to exhibit quantitative agreement with realistic three dimensional models  \cite{Majorosi2018,Majorosi2020}.  

The external laser field is either taken to be static, that is $E(t)=E_0$,  or a time-dependent linearly polarised polychromatic field such that
\begin{equation}\label{eq:polychrom}
    E(t)=E_0 \left[ \mathrm{cos}(\omega t) + r_t \mathrm{cos}(b \omega t+\phi) \right],
\end{equation}
where $b$ is the frequency ratio between the first and the second driving wave, $r_t$ is the field-strength ratio, $\phi$ is the relative phase between the two driving waves, and $\omega$ is the frequency of the fundamental.

\begin{figure}
    \centering
    \includegraphics[width=0.6\linewidth]{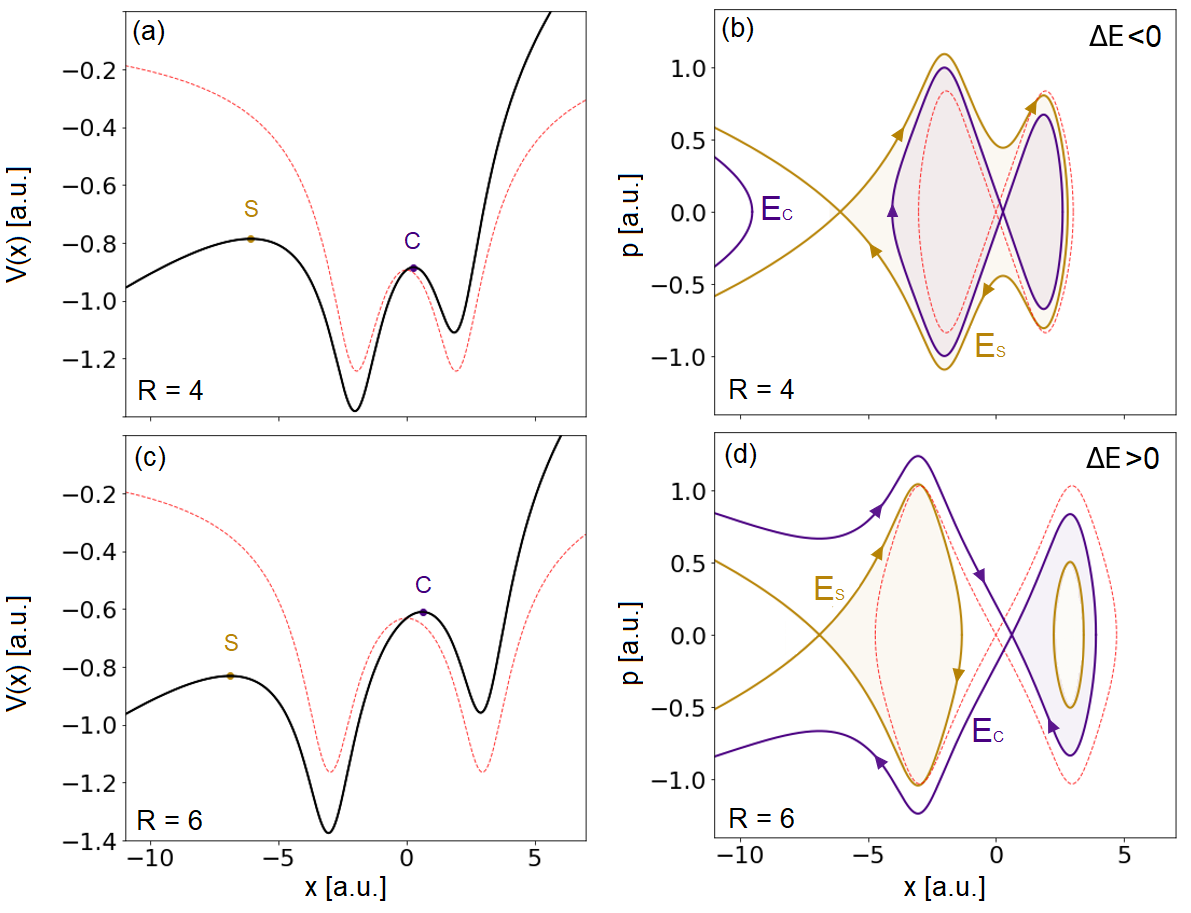}
    \caption{Phase portraits for the one-dimensional homonuclear molecular models described by the binding potentials (\ref{eq:potential}), using internuclear separations of $R = 4$ and $R = 6$ and a static field $E=0.07.$ a.u. [left panels], together with the corresponding effective potentials [right panels]. The Stark and the central saddles are indicated by the labels S and C in the figure, and their energy by the labels $\mathrm{E_S}$ and $\mathrm{E_C}$. The field-free separatrices and potentials are given by the dashed red lines.
    }
    \label{fig:phaseSpace}
\end{figure}

This molecular model is analysed in phase space. There, the bound and continuum regions may be identified, and an example of is shown in Figure~\ref{fig:phaseSpace}. Fixed points (points that satisfy $\dot{x}=\dot{p}=0$) here are either centres or saddles \cite{Arrowsmith1992}. Centres correspond to minima of the effective potential, and are surrounded by closed orbits. Saddles correspond to maxima of the effective potential, and delimit different dynamical regions of phase space. For instance, the central saddle separates the two molecular wells with energy $E_C$, and the Stark saddle is located between the downfield well and the continuum with energy $E_S$. The energy curves of those saddles in phase-space are called separatrices. We define the energy difference between these two saddles as
\begin{equation}
    \label{eq:sepEnergyDiff}
   \Delta E = E_C - E_S.
\end{equation}
$\Delta E$ is either negative (Fig.~\ref{fig:phaseSpace}(a)) or positive (Fig.~\ref{fig:phaseSpace}(c)) and characterises two different phase space configurations: closed (Fig.~\ref{fig:phaseSpace}(b)) and open (Fig.~\ref{fig:phaseSpace}(d)) separatrices. 

The initial state of the electron $\Psi(x, 0)$ is approximated as a Gaussian wavepacket,
\begin{equation}\label{eq:gaussian_i}
    \Psi(x, 0)=\langle x \mid \Psi(0)\rangle=\left(\frac{\gamma}{\pi}\right)^{\frac{1}{4}} \exp \left\{-\frac{\gamma}{2}\left(x-q_{0}\right)^{2}+\mathrm{i} p_{0}\left(x-q_{0}\right)\right\},
\end{equation}
of width $\gamma$ and initial momentum $p_0 = 0$. The state shall be called localised in the left well when $q_0 = - R/2$, denoted $\Psi_{\mathrm{l}}(x, 0)$, or in the right well when $q_0 = R/2$, denoted $\Psi_{\mathrm{r}}(x, 0)$. Alternatively, we may form delocalised coherent superpositions. 
We introduce some asymmetry between the initial left and right wavepackets in a delocalised setting. As such, the initial state in \cite{chomet2019quantum} is generalised by including a relative phase $\theta \in [-\pi, \pi]$ and a localisation parameter $\alpha \in [0, 1]$, giving
\begin{equation}\label{eq:genwp}
    \Psi_{\alpha, \theta}(x, 0) \coloneqq \frac{\sqrt{\alpha} \Psi_{\mathrm{l}}(x, 0)+\sqrt{1-\alpha} e^{i \theta} \Psi_{\mathrm{r}}(x, 0)}{\sqrt{1+2 \sqrt{\alpha(1-\alpha)} \cos (\theta) \mathcal{I}_{o}}}.
\end{equation}
Under some circumstances, we also consider different widths $\gamma_l$ and $\gamma_r$ for the wavefunctions $\Psi_{\mathrm{l}}$ and $\Psi_{\mathrm{r}}$, respectively. 
The constant of normalisation in the denominator depends on the initial state overlap $\mathcal{I}_0$:
\begin{equation}
    \mathcal{I}_{0}=\int \Psi_{\text {l}}^{*}(x, 0) \Psi_{\text {r}}(x, 0) \mathrm{d} x. 
\end{equation}

Note that $\Psi_{1, \theta}(x, 0) = \Psi_{\mathrm{l}}(x, 0)$ and $\Psi_{0, \theta}(x, 0) = \Psi_{\mathrm{r}}(x, 0)$ (up to a global phase), whilst intermediate values of $\alpha$ produce coherent superpositions of varying weight. 
If one considers $\Psi_{\frac{1}{2}, 0}(x, 0)$ and $\Psi_{\frac{1}{2}, \pi}(x, 0)$, this will lead to even and odd cat states, for which the wavepacket localisation in each centre are equally weighted. 

\subsection{Relevant quantities}
\label{sec:tools}
Of particular interest is the phenomenon of \textit{enhanced ionisation}, characterised by anomalous peaks in the ionisation rate for relatively large internuclear separations. The effect is understood to arise from the presence of strongly coupled charge resonant states in combination with a narrowing of the effective potential for the upfield charge centre \cite{zuo1995charge,Seideman1995}. The latter is a result of the neighbouring downfield centre enabling efficient population transfer from upfield to the continuum. The ionisation rate $\Gamma$ from time $t=0$ to $t=T$ is quantified using
\begin{equation}
\label{eq:rate}
\Gamma=-\ln \left(\frac{|\mathcal{P}(T)|^{2}}{|\mathcal{P}(0)|^{2}}\right) \frac{1}{T}
,\end{equation}
where $\mathcal{P}(t)$ is the probability function at $t$, given by
\begin{equation}
\mathcal{P}(t)=\int^{+\infty}_{-\infty}\Psi^*(x,t)\Psi(x,t)dx.
\label{eq:prob}
\end{equation} 
This definition of ionisation rate was used in the seminal paper \cite{zuo1995charge} in the context of enhanced ionisation of molecules. Numerically, the integral limits of Eq.~(\ref{eq:prob}) are finite, taken throughout the box of size \(x=-100\) to \(x=100\) a.u. Because of irreversible ionisation Eq.~(\ref{eq:prob}) will decrease with time and be less than unity. No absorber was used, and reflections are minimised by using a total grid size twice as large as the box size. The final time is chosen to be $T = 150~\mathrm{ a.u.}$ as it has been used in previous publications \cite{chomet2021attoscience,chomet2019quantum,kufel2020alternative}. Throughout, $\Gamma$ will be written in arbitrary units. 

Eq.~(\ref{eq:rate}) will be employed throughout in order to test our optimisation methods for static fields. The static field in this work is chosen in such a way that the left well is located downfield and the right well upfield.

Another relevant quantity, which provides insight into the resulting wave function dynamics, is the autocorrelation function 
\begin{equation}\label{eq:auto}
    a(t)=\int \Psi^{*}(x, t) \Psi(x, 0) \mathrm{d} x,
\end{equation}
which is given by the overlap integral between the initial and time propagated wave function. The autocorrelation function will be employed to assess the behaviour of the system in time-dependent electric fields. 

The Wigner quasi-probability distribution  \cite{Wigner1932} is a useful tool in the analysis of phase-space dynamics (for a review see \cite{Weinbub2018}). It is defined as
\begin{equation}\label{eq:wigner}
W(x, p, t)=\frac{1}{\pi} \int_{\infty}^{-\infty} \mathrm{d} \xi \Psi^{*}(x+\xi, t) \Psi(x-\xi, t) \mathrm{e}^{2 \mathrm{i} p \xi}.
\end{equation}
The Wigner function is real and normalised, with the property that its marginals correspond to physical probability distributions for each conjugate variable, respectively. W(x,p,t) may assume negative values, and this can provide an indication of nonclassicality \cite{benedict1999wigner}.

\section{Dimensionality Reduction}
\label{sec:dimReduction}

The present problem depends on many parameters: The initial wavepacket localisation $\alpha$, its widths $\gamma_l$ and $\gamma_r$ and relative phase $\theta$; molecular parameters like the weights $Z_l$ and $Z_r$ and the inter-nuclear distance $R$; and many external electric-field parameters such as $\phi$, $r_t$, $b$ and $E_0$. The main challenge to overcome when studying the effect of a large number of parameters simultaneously is how to visualise the results. Indeed, each data point will exist in a high-dimensional space (the number of dimensions equal to the number of parameter used), and ideally we would like to project our results down to a two-dimensional space.

To this end we will use T-Distributed Stochastic Neighbour Embedding (t-SNE), an unsupervised dimensionality reduction technique, meaning it will find overall trends and patterns without any prior knowledge on the origin of the data. It can be briefly summarised in two steps:

\begin{itemize}
    \item Define $G_{xy}$ and $C_{xy}$ as the probability of picking a pair of points (x,y) in high dimensional and two dimensional space respectively. Both probabilities are characterised such that they are larger if the neighbouring points are close. $G_{xy}$ is a Gaussian distribution and $C_{xy}$ a Cauchy distribution.
    \item Minimise the Kullback–Leibler divergence \cite{Kullback1951} between the two distributions with respect to their projection in two-dimensional space using the gradient descent method \cite{Cauchy1847}.
\end{itemize}

A more detailed explanation of the workings of the t-SNE method can be found in \cite{vandermaaten08a:tsne_original_paper}. As a reality check, we will compare all t-SNE projections to Principal Component Analysis (PCA) \cite{Pearson1901} dimensional reduction. Indeed, t-SNE is stochastic, so each run can yield different results. It focuses on preserving the local structure of the data and is a non-linear technique. On the other hand, PCA is linear, deterministic, and preserves global properties while potentially losing low-variance deviations between neighbours. Therefore by comparing both projections we can guarantee the accuracy of the t-SNE results. For clarity, some key results obtained with the PCA are given in the Appendix A.
To make use of the t-SNE capabilities we need to include information about the wavefunction in the static field case or its evolution in the time-dependent case. 

\subsection{Time evolution characterisation}
\label{sec:characterisation}

\begin{figure}[tbp]
    \centering
    \includegraphics[width=\textwidth]{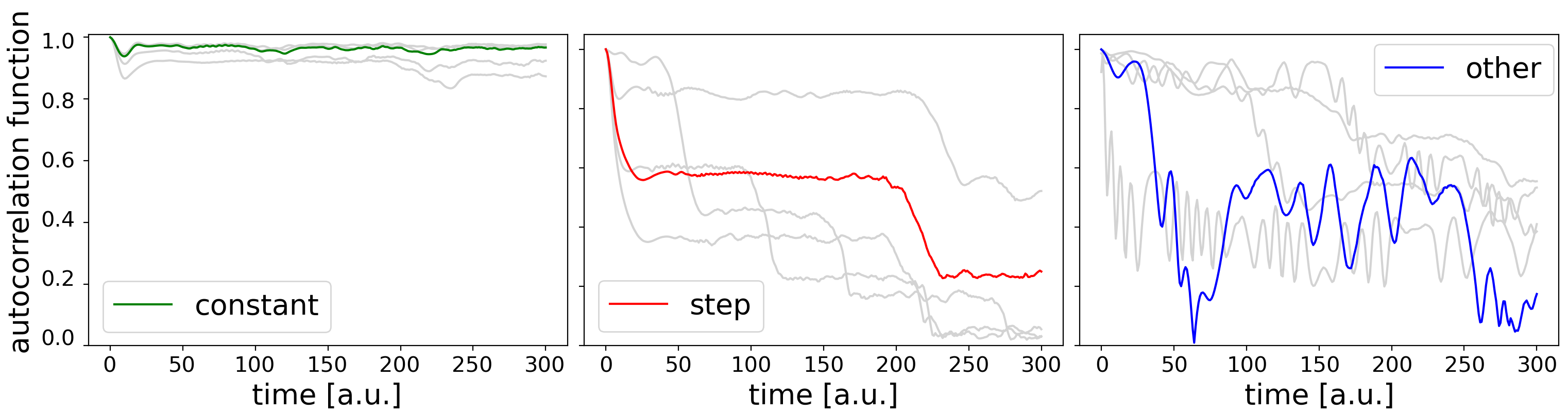}
    \caption{Examples of the three potential outputs of the autocorrelation function sorting algorithm. The parameters used vary greatly but are all within the range shown in Table.~\ref{table:paramTable}.}
    \label{fig:functSorting}
\end{figure}

In order to understand the system dynamics through the use the dimensionality reduction methods, we wish to characterise the electronic dynamics by a single value. In the static-field case we will employ the ionisation rate $\Gamma$ (\ref{eq:rate}). For the time-dependent case we choose to quantify the shape of the autocorrelation function (\ref{eq:auto}), as it carries more temporal information. A preliminary investigation showed that the autocorrelation function can take three distinct shapes (Figure \ref{fig:functSorting}): `steps', `constant', and `other', which represent different behaviours.

The classification of autocorrelation functions into the defined categories is performed by a simple deterministic decision tree algorithm. First, we check whether a function is `constant', that is, if it does not fall below a given threshold value. Next, we look for the steps. Steps are characterised by the intervals where the autocorrelation function stays approximately constant, followed by a steep decline. Hence, if we find a constant interval longer than some `critical length', we categorise it as a `step'. We define a constant interval as an interval where the difference of any two values is smaller than a critical threshold, `$\epsilon$'. If the function is not caught by these checks, it is classified as `other'.
In the results presented in section \ref{sec:TD} the threshold value is set to 0.8, the critical length is set to 100, while the critical $\epsilon$ is 0.03. All of the autocorrelation function arrays used have 3000 elements. 

\subsection{Method}
\label{sec:dataRed}
To obtain a data point, $N-1$ parameters are randomly uniformly distributed within the range shown in Table.~\ref{table:paramTable}.
The electronic wavefunction is then evolved using the time-dependent Schr\"odinger equation and we characterise the time evolution information by a single value as described in the above section In case of a static external field studied in section \ref{sec:static} this is done by computing the ionisation rate $\Gamma$ (Eq.~(\ref{eq:rate})). For the time-dependent field in section \ref{sec:TD} we quantify the shape of the electron autocorrelation function (see section \ref{sec:characterisation}). The $N-1$ parameters chosen randomly as well as the single value characterising the time evolution form a data point of size $N$.

In certain situations specific data points are removed from the final data set in order to focus the visualisation on the matter at hand. For example in section \ref{sec:TDtsneResults}, only 1332 data points are used from the 1 million originally computed using random parameters. Indeed, in order to focus on the `step' output of the autocorrelation function, approximately 99.9\% of data points with outputs `constant' and 99.97\% of data points with output `other' are removed.  

\begin{figure}
    \centering
    \includegraphics[width=0.9\linewidth]{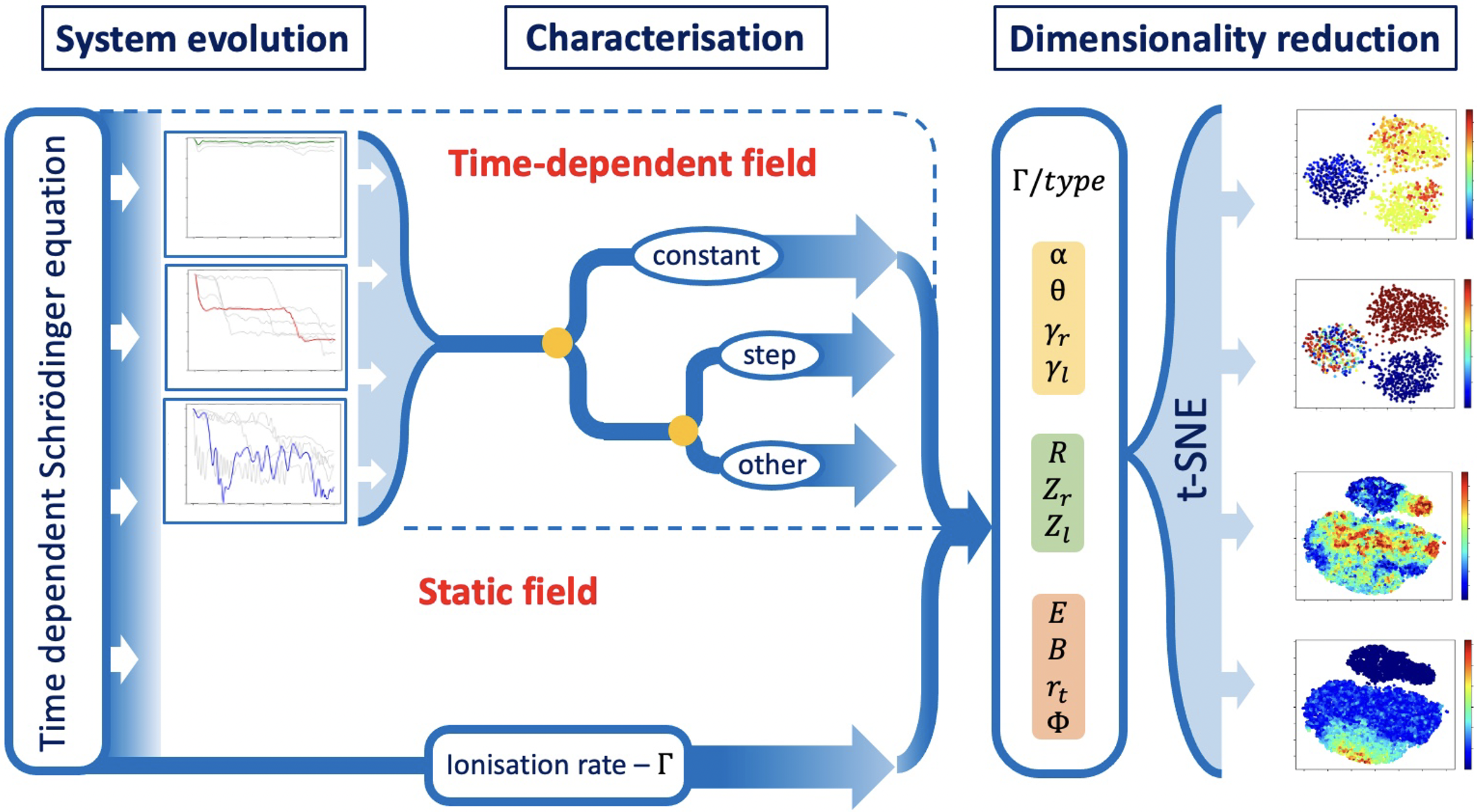}
    \caption{Schematic outline of the methodology used throughout the paper.}
    \label{fig:schematic}
\end{figure}

We then use the t-SNE to reduce our N-dimensional data points to 2 dimensions, projecting the data set onto the two axis $\mathrm{y}_1$ and $\mathrm{y}_2$. The data points will have clustered into different groups, the nature of which will be understood by plotting the projection as a function of one of the input parameters. This will allow us to determine which parameter or which combination of parameters influence the ionisation rate or the autocorrelation function step, and to what degree. The overall method described here and used throughout this paper is summarised schematically in Fig.~\ref{fig:schematic}. Finally, the PCA method is used to ensure that the observed behaviour is neither a remnant of the stochastic nature of the algorithm, nor due to the removing of data points from the data set. Examples of PCA results are shown in Appendix A.

\begin{table}
%label{table:paramTable}
\begin{center}
\begin{tabular}{||c| c c c c c c c c ||} 
\hline
  Static Field &$\alpha$ & $\theta$ & $\gamma_r$ & $\gamma_l$ & R & $Z_r$ & $Z_l$ & $E_0$\\ [0.5ex] 
 \hline\hline
 min & 0.0 & 0.0 & 0.2 & 0.2 & 3.0 & 1.0 & 1.0 & 0.04\\ 
 \hline
 max & 1.0 & $\pi$ & 1.0 & 1.0 & 12.0 & 2.0 & 2.0 & 0.08\\
 \hline

\end{tabular}
\begin{tabular}{||c| c c c c c c c c c c c||} 
\hline
  Time-dependent Field &$\alpha$ & $\theta$ & $\gamma_r$ & $\gamma_l$ & R & $Z_r$ & $Z_l$ & $E_0$ & $b$ & $r_t$ & $\phi$\\ [0.5ex] 
 \hline\hline
 min & 0.0 & 0.0 & 0.2 & 0.2 & 4.0 & 1.0 & 1.0 & 0.04 & 0.5 & 0.0 & -$\pi$\\ 
 \hline
 max & 1.0 & $\pi$ & 1.0 & 1.0 & 10.0 & 1.5 & 1.5 & 0.08 & 2.0 & 1.0 & $\pi$ \\
 \hline
\end{tabular}
\end{center}
\caption{Table of all parameters used throughout this paper with the minimum and maximum value of the range they are randomly chosen from.\label{table:paramTable}
}
\end{table}

\section{Static fields}
\label{sec:static}

We will start our study with an analysis of enhanced ionisation in static fields. While the ultimate goal is to understand and control enhanced ionisation in time-dependent fields, the purpose of this static field section is two-fold:

\begin{itemize}
    \item Obtain an in depth understanding of how certain parameters influence the system, and expand on the conclusions drawn in \cite{chomet2021attoscience,chomet2019quantum}.
    \item Establish the effectiveness of the t-SNE dimensionality reduction technique as a method of simultaneous multi-parameter analysis. This will allow us to use it as a primary tool in section \ref{sec:TD}.
\end{itemize}

\subsection{Qualitative analysis}
\label{sec:qualitative}
As a starting point, we will perform a qualitative investigation of how specific parameters influence the quantum bridges and the ionisation rate in a static field using the generalised initial wavepacket given by Eq.~\eqref{eq:gaussian_i}. We employ phase-space arguments and Wigner functions to facilitate the interpretation.

\subsubsection{Initial electron wavepacket.}
\label{sec:initialwp}

\begin{figure}[ht]
    \centering
    \includegraphics[width=0.5\textwidth]{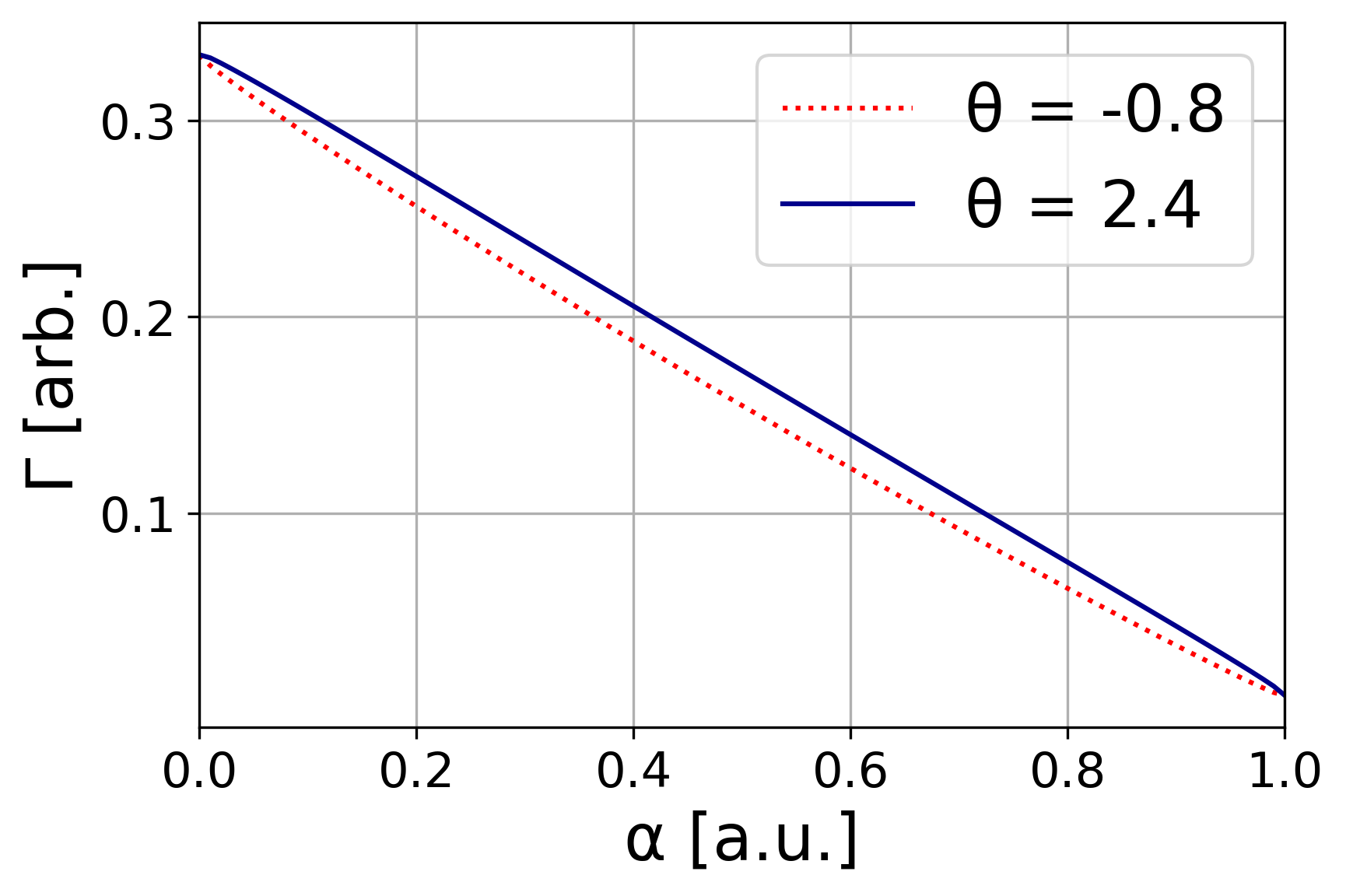}
    \caption{Ionisation rate $\Gamma$ for an initial wave packet $\Psi_{\alpha, \theta}(x, 0)$ with respect to the localisation parameter $\alpha$, comparing maximum $\theta = 2.4$ and minimum $\theta = -0.8$. The external static field has strength $E_0 = 0.06$, and the molecular parameter used are internuclear distance $R = 7$, $Z_r = 1.0$ and $Z_l = 1.0$.}
    \label{fig:relAmpl}
\end{figure}

Our earlier work on the subject \cite{chomet2019quantum} led to the conclusion that a major predictor of enhanced ionisation was the localisation of the initial electron wavepacket. However, only three configurations were studied: localised around the right well, localised around the left well, and delocalised equally around both centres. Here we start with a more thorough analysis of this parameter by considering it a continuous variable, $\alpha$, see Eq.~(\ref{eq:genwp}). As shown in Fig.~\ref{fig:relAmpl}, the ionisation rate is linearly proportional to the electron localisation, with $\alpha = 0.0$ (initially localised upfield) leading to the highest ionisation rate. This can be understood as the downfield centre narrowing the effective-potential barrier for the upfield centre, hence enhancing upfield ionisation.

In addition to a continuous localisation parameter, the state of the initial electron wavepacket is varied by a relative phase factor $\theta$, see Eq.~(\ref{eq:genwp}), between the superposition of $\Psi_r$ and $\Psi_l$. When using an equally delocalised wavepacket, $\alpha = 0.5$, the ionisation rate as a function of $\theta$ (in rad) peaks around $\theta = 2.4$ and is minimal for $\theta=-0.8$. This has little overall influence on the ionisation rate as seen in Fig.~\ref{fig:relAmpl}. Finally, the initial wavepacket widths $\gamma_r$ and $\gamma_l$, see Eq.~(\ref{eq:gaussian_i}), are set to differ from $\gamma = 0.5$, which is the minimal ground-state energy of a field free single-centre soft-core potential with softening parameter $a = 1.0$. This has no major influence on the ionisation rate and is not shown here.

\subsubsection{Heteronuclear molecule $Z_{l,r}$.}

\begin{figure}
    \centering
    \includegraphics[width=0.6\linewidth]{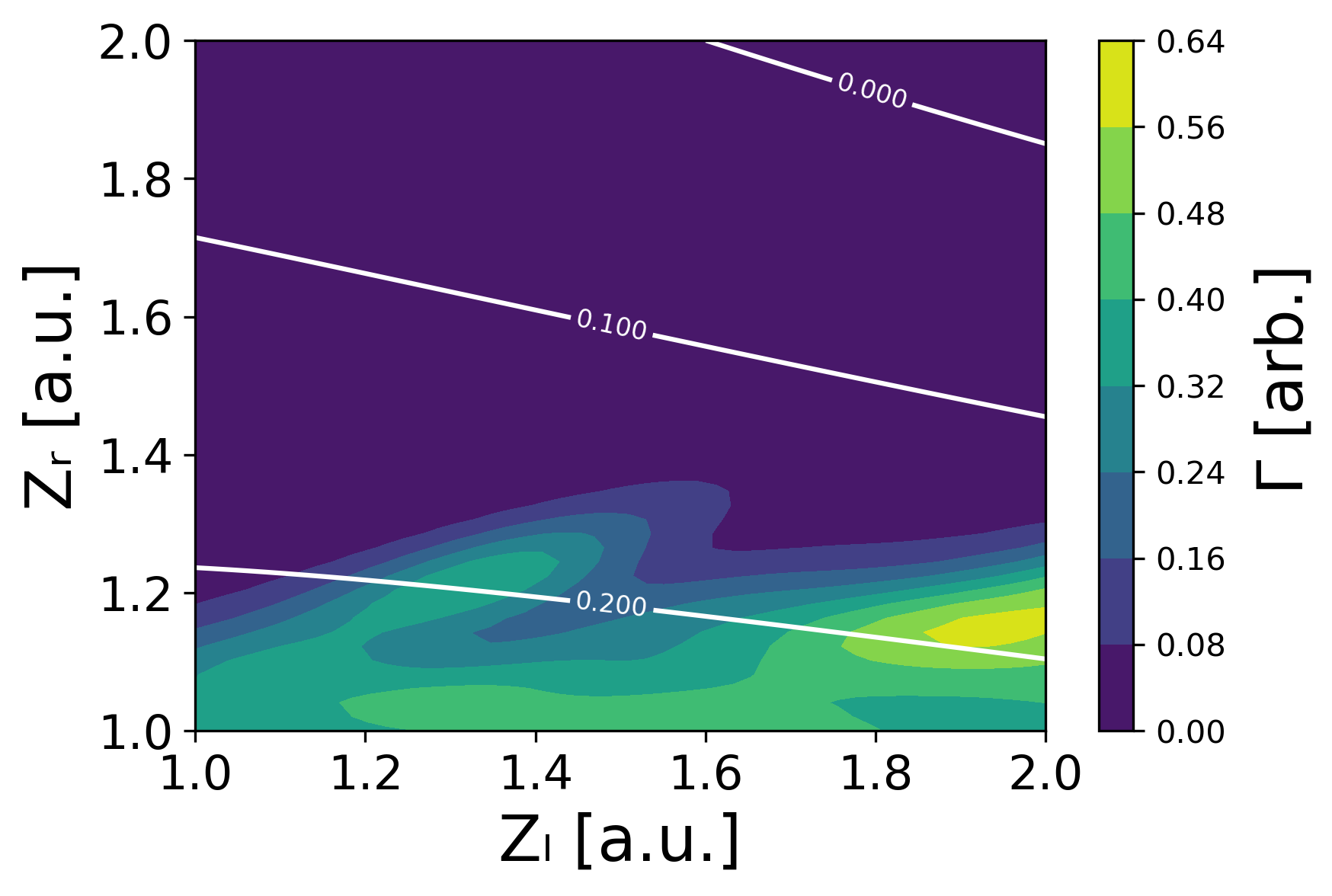}
    \caption{Contour plot of the ionisation rate $\Gamma$ as a function of $Z_l$ and $Z_r$ using an external static field $E_0 = 0.06$ with internuclear distance $R = 7$. The white lines indicate the separatrix energy difference $\Delta E$, with the 0.0 line indicating the transition from open separatrices ($\Delta E$ > 0) to closed separatrices ($\Delta E$ < 0).}
    \label{fig:rate-contour}
\end{figure}

\begin{figure}
    \centering
    \includegraphics[width=\linewidth]{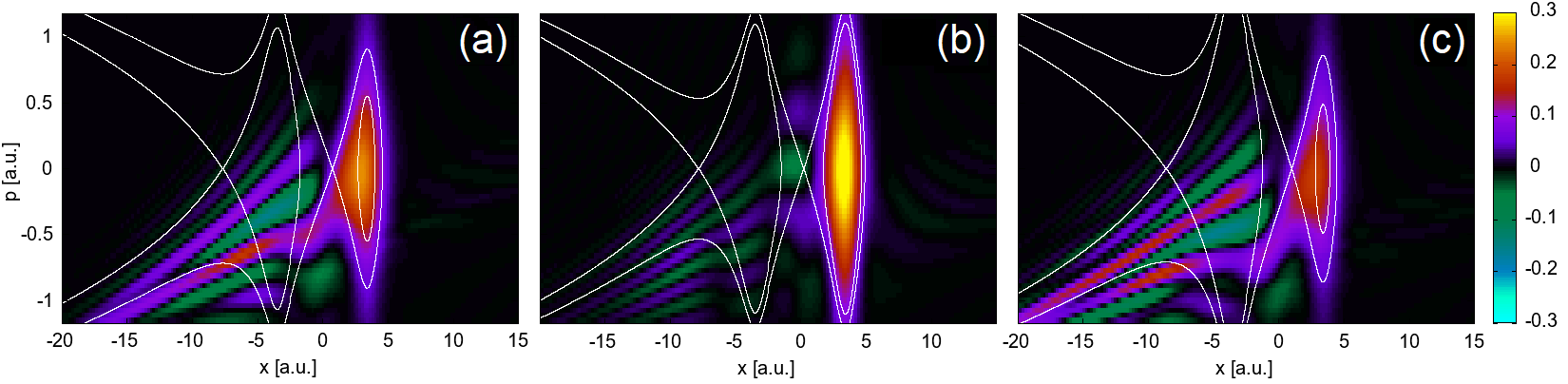}
    \caption{Wigner quasi-probability distribution of the ionisation from a static field $E_0 = 0.06$ with internuclear distance $R = 7$ for (a) $Z_r = 1.0$ and $Z_l = 1.0$, (b) $Z_r = 1.5$ and $Z_l = 1.0$ and (c) $Z_r = 1.0$ and $Z_l = 1.5$. The snapshots are taken at $t = 25~\mathrm{a.u.}$.}
    \label{fig:WignerZrZl}
\end{figure}

Finally, we expand our study to include heteronuclear molecules. We analyse the ionisation rate while varying the value of molecular weights $Z_r$ and $Z_l$, of Eq.~\ref{eq:potential}. The study of heteronuclear molecules is motivated by enabling a change in the critical internuclear distance where the phase-space separatrices change configuration and in the dynamics of the quantum bridges. From Fig.~\ref{fig:rate-contour} studying the ionisation rate as a function of both charges $Z_r$ and $Z_l$ simultaneously shows a large region of suppressed ionisation as  $Z_r > 1.3$. The effect of $Z_l$ is a lot less pronounced, leading to high ionisation rates even at $Z_l = 2.0$. While open separatrices do not guarantee a high ionisation rate, closed separatrices always suppress ionisation. To have a better understanding of the sharp decrease in ionisation rate with respect to $Z_r$, we look at the Wigner quasi-probability distribution in Fig.~\ref{fig:WignerZrZl}. Comparing the bound region around the right molecular centre in (a) and (b), as $Z_r$ increases, the right molecular well deepens greatly. This leads the upfield population to stay trapped and greatly suppresses ionisation. On the other hand, the increase of the left bound region in (c) due to an increase in $Z_l$ does not greatly influence the ionisation rate. This is because in enhanced ionisation the majority of the ionised population comes from the upfield centre, as shown in \cite{chomet2019quantum}. There is very little change to the upfield (right-hand-side here) centre when increasing $Z_l$. 

Now, a more complete overview would go through the effect of $Z_l$ and $Z_r$ with respect to various parameters, which would be slow and exhaustive. Instead we propose in the following section to use dimensionality reduction techniques to analyse the effect of multiple parameters at once. 

%%%%%%%%%%%%%%%%%%%%%%%%%%
\subsection{Proof of concept - machine learning techniques}
\label{sec:staticTSNE}

We now aim to study the effect of multiple parameters on the ionisation rate, starting with a limited number in order to fully understand the results and to focus on the effect of the chosen parameters. Those are the internuclear distance R, the width of the initial superposed gaussian upfield and downfield wavepackets $\gamma_l$ and $\gamma_r$, the phase difference $\theta$ between them, and the localisation parameter of the wavepacket $\alpha$. The value of these parameters is chosen from a predetermined range described in section \ref{sec:dataRed}.

\begin{figure*}[tbp]
\centering
\includegraphics[width=\textwidth]{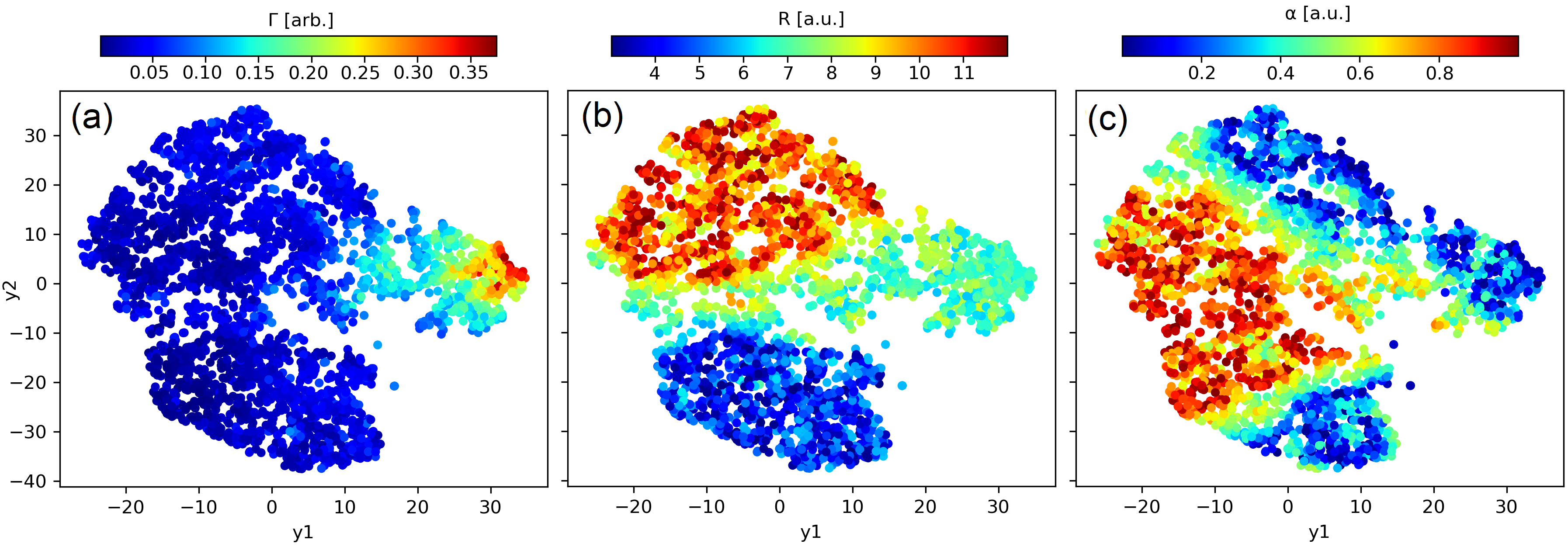}
\caption{Plot of the 3000 6-dimensional data points that have been projected onto two axis $\mathrm{y}_1$ and $\mathrm{y}_2$ using the t-SNE dimensional reduction technique described in section \ref{sec:tools}. Each subplot has the projected data plotted as a function an initial input parameter. The initial input parameters include (a) the ionisation rate $\Gamma$ (given in arbitrary units), (b) the internuclear distance $R$, (c) the electron localisation $\alpha$. Also included but not shown here are the left initial wavepacket width $\gamma_l$, the right wavepacket width $\gamma_r$ and the phase difference between the left and right initial wavepackets $\theta$. PCA distributions using the same data points have been analysed for accuracy (not shown). The external field is taken to be static with $E_0 = 0.0534$ and the molecular weights $Z_r = 1.0$ and $Z_l = 1.0$, matching the results from \cite{chomet2019quantum}. }
\label{fig:tsne1}
\end{figure*}

Data points comprised of these five parameters as well as the ionisation rate are projected onto two axis $\mathrm{y}_1$ and $\mathrm{y}_2$ using the t-SNE algorithm, shown in Figure~\ref{fig:tsne1}. This method is described in section \ref{sec:dataRed}.%, and the results are compared with a PCA projection.

From Figure~\ref{fig:tsne1}, we see that the projected data bring into focus the connection between the ionisation rate, the internuclear distance and the electron localisation. These match the results from \cite{chomet2019quantum} and section \ref{sec:qualitative}. That is, peak ionisation yield is present at and around the peak internuclear distance of $R = 6.8$ a.u., shown in Figure~\ref{fig:tsne1}(b), and when the initial wavepacket is localised upfield, shown in Figure~\ref{fig:tsne1}(c). 
The low ionisation rate regions, seen in Figure~\ref{fig:tsne1}(a), split into two distinct parts seen in Fig.~\ref{fig:tsne1}(b) which correspond respectively to regions of too low and too high internuclear distance, studied in depth in \cite{chomet2019quantum}. The ionisation rate falls linearly as the electron localisation changes from localised upfield to localised downfield, as seen in Fig.~\ref{fig:relAmpl}. The effect of the gaussian widths $\gamma$ and the phase difference $\theta$ is a lot weaker than those discussed above, and are not shown here.

We will now look at the full range of parameters. In addition to the previous six, these following results will include the effects of the nuclear weight number $Z_r$ and $Z_l$, and the electric field strength $E_0$. Because of these additional parameters, we posit that there is no longer a clear internuclear distance peak, but instead a larger range of optimal internuclear distances, dependent on $E_0$, $Z_r$ and $Z_l$. For that reason  the difference in energy between the central and Stark separatrix $\Delta E$ will also be analysed, but will not be an input parameter. 

\begin{figure*}[tbp]
\begin{minipage}{0.49\textwidth}
\centering
\includegraphics[width=\textwidth]{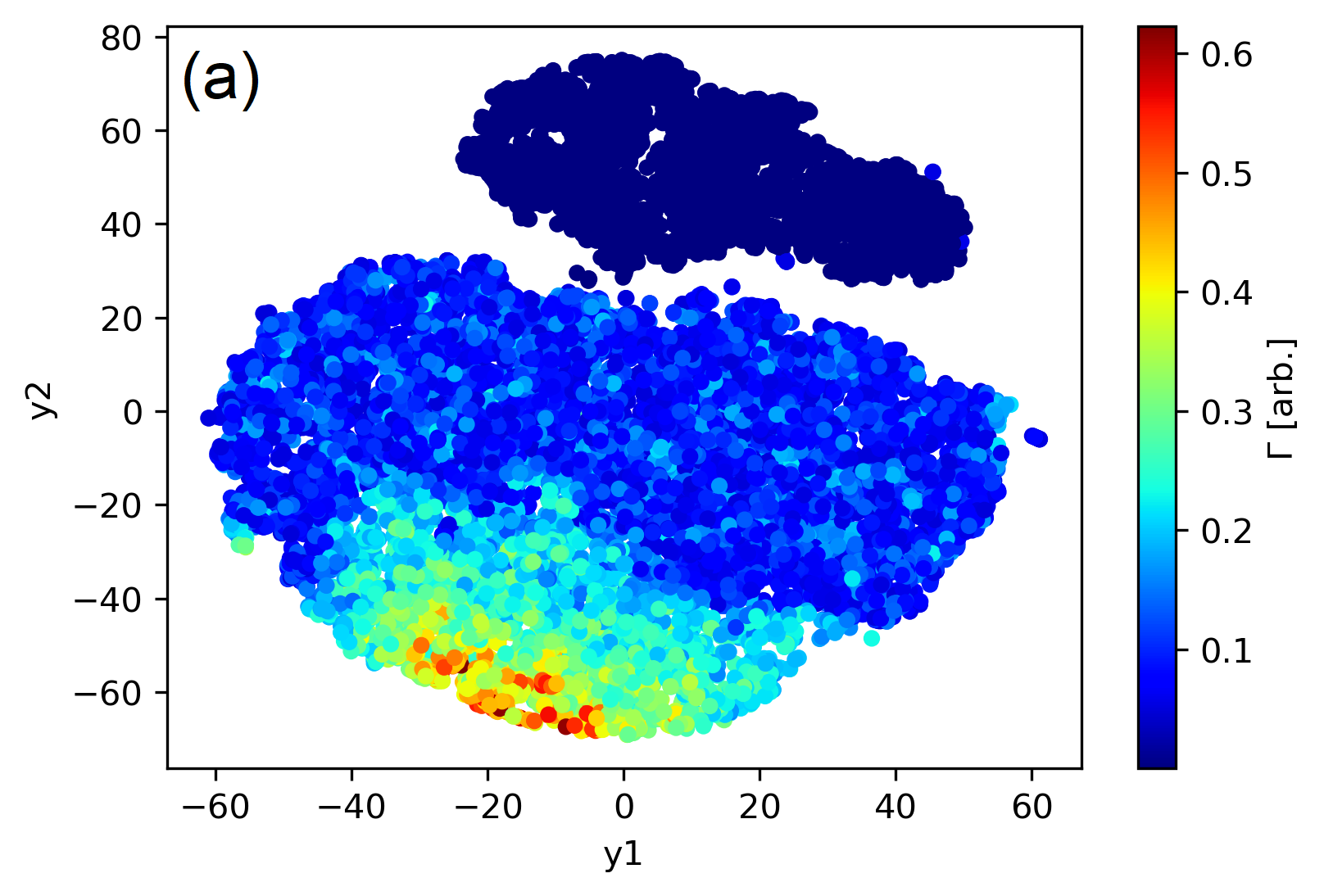}
\end{minipage}
\begin{minipage}{0.49\textwidth}
\centering
\includegraphics[width=\textwidth]{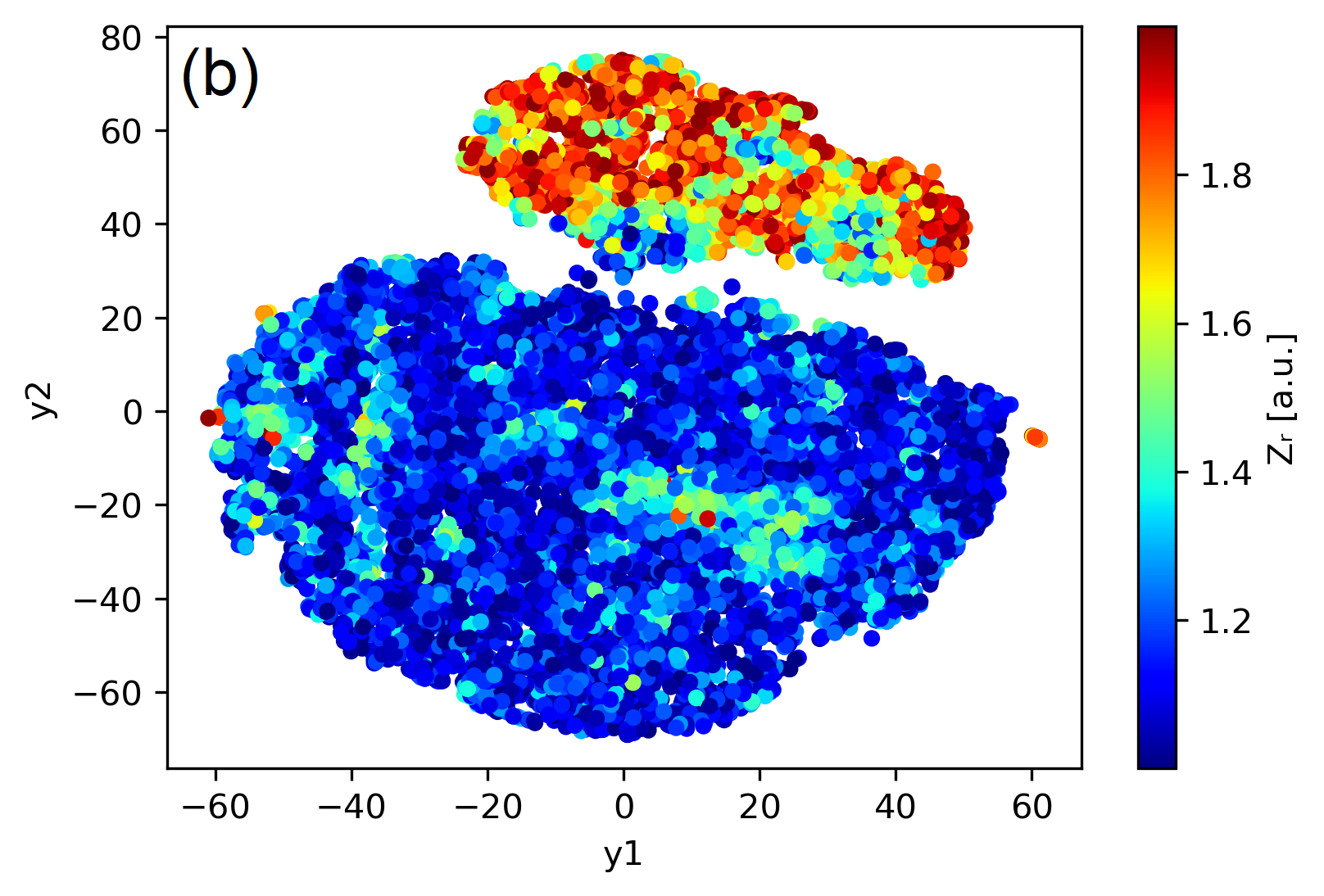}
\end{minipage}\\
\begin{minipage}{0.49\textwidth}
\centering
\includegraphics[width=\textwidth]{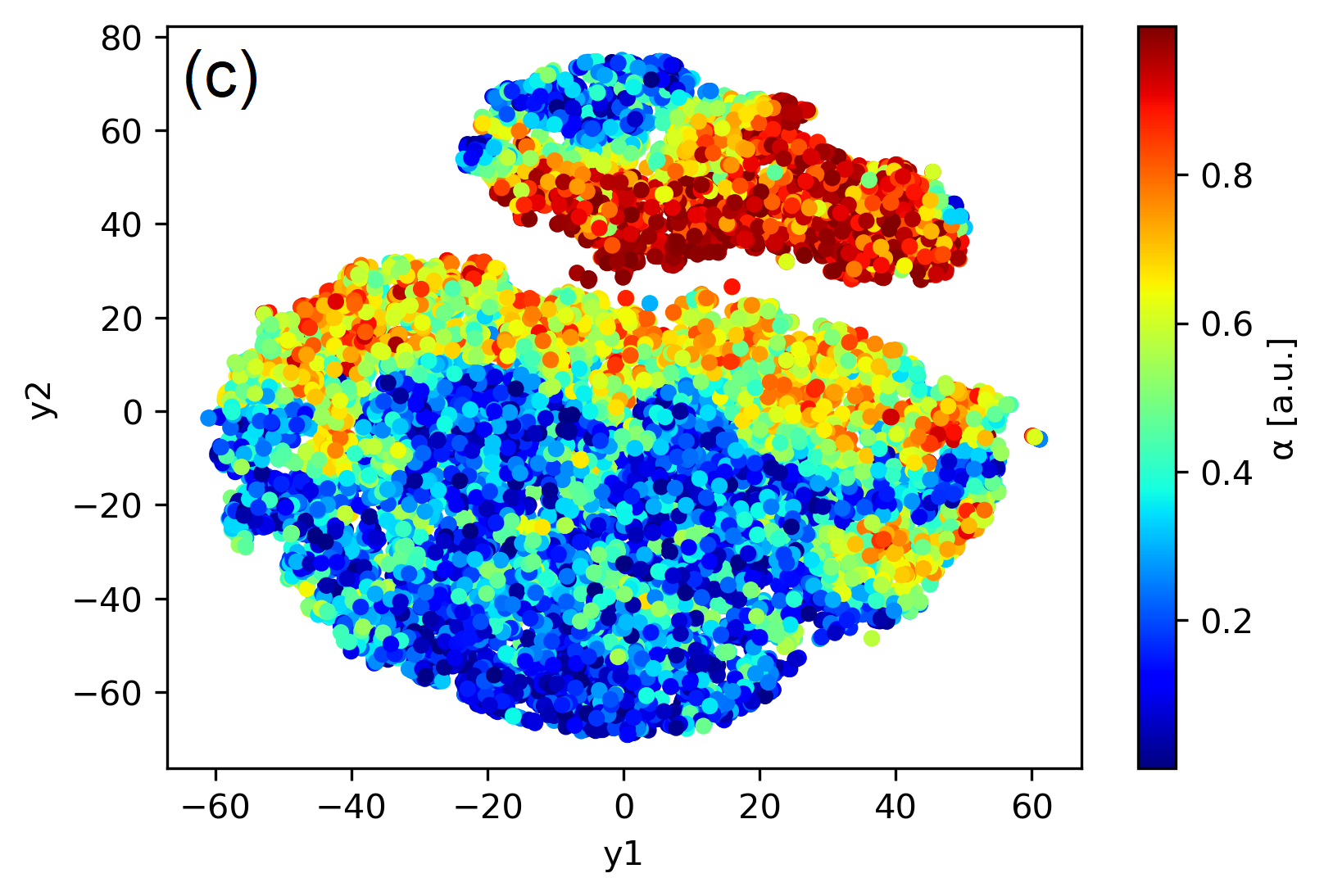}
\end{minipage}
\begin{minipage}{0.49\textwidth}
\centering
\includegraphics[width=\textwidth]{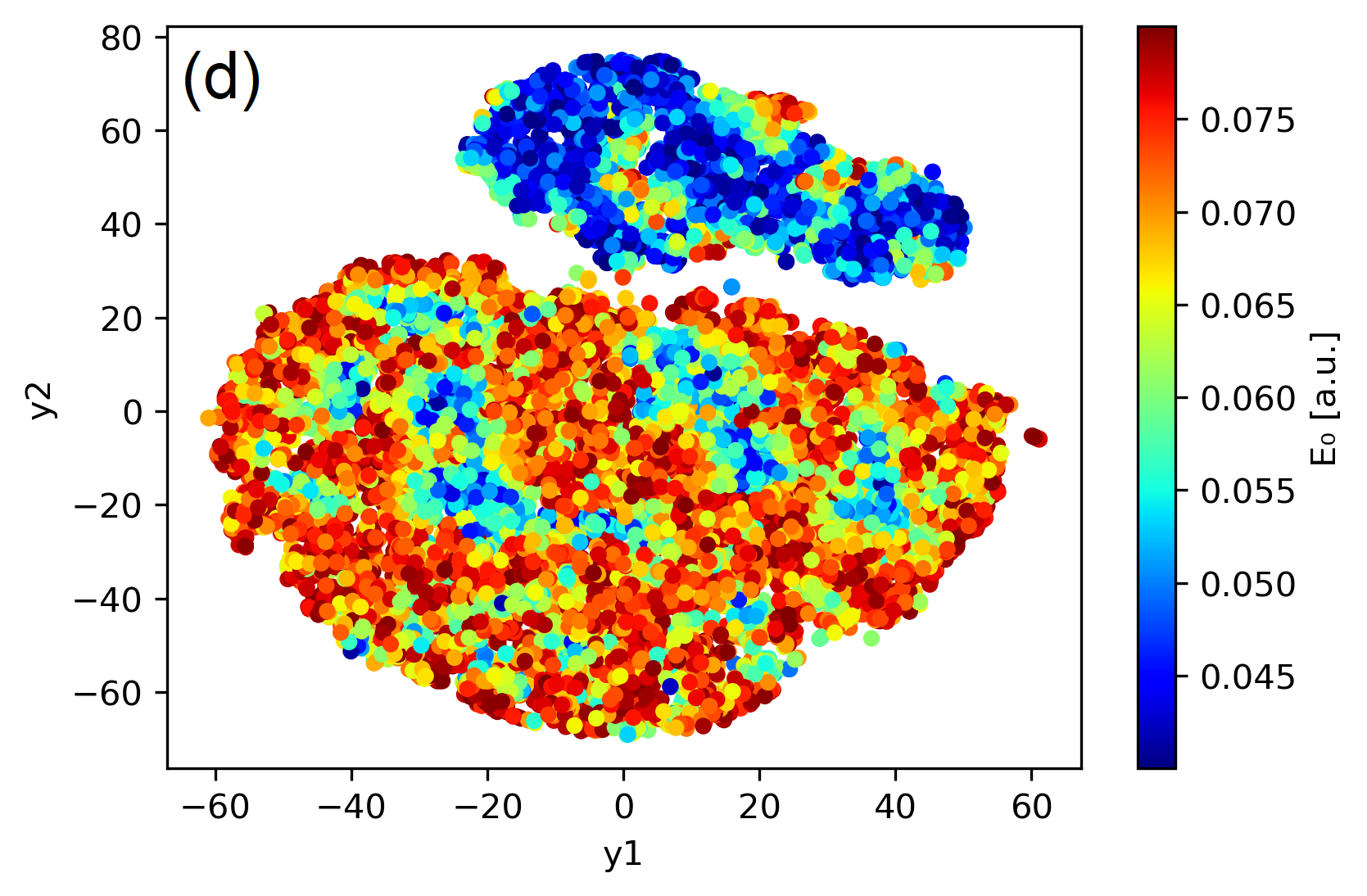}
\end{minipage}\\
\begin{minipage}{0.49\textwidth}
\centering
\includegraphics[width=\textwidth]{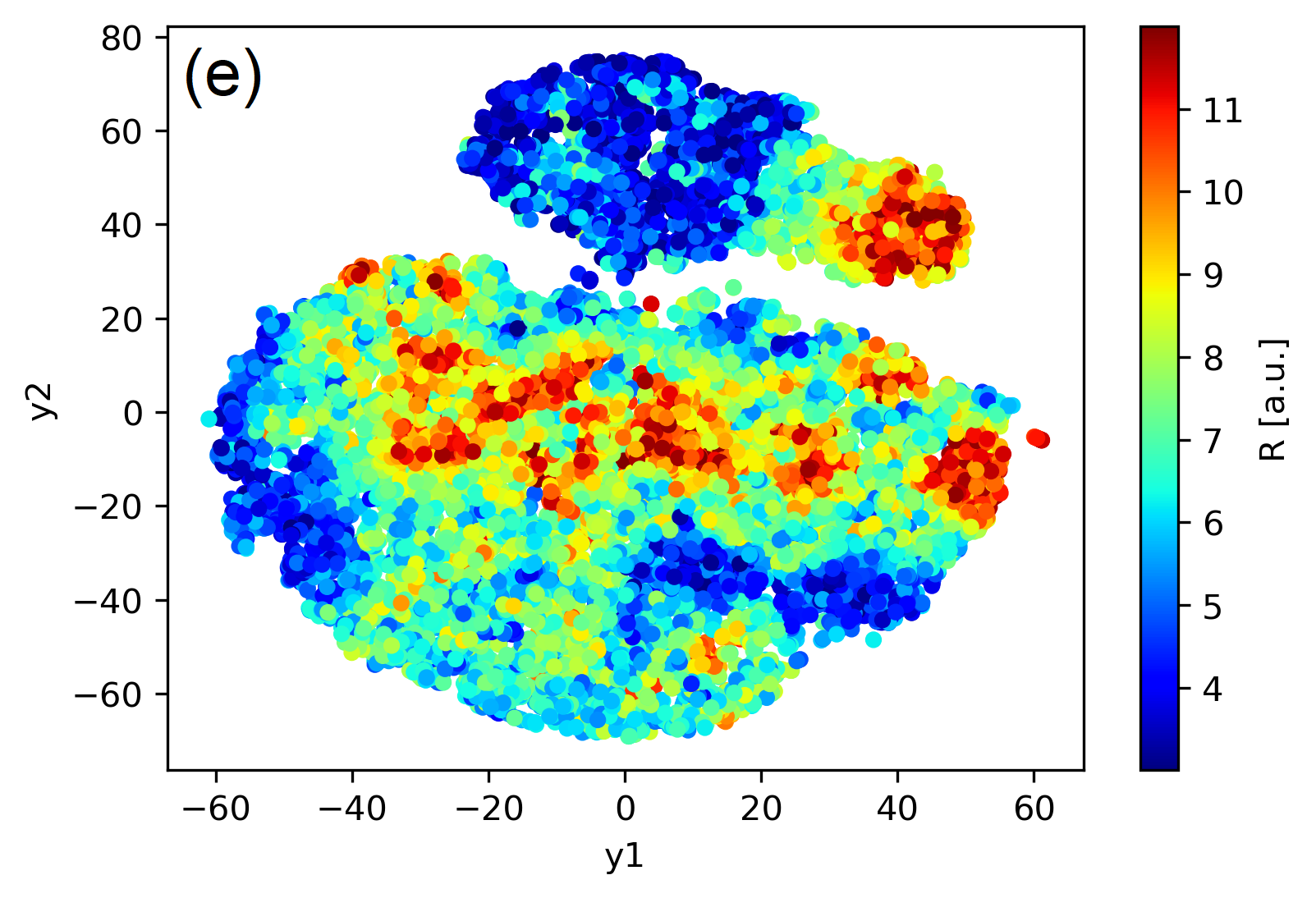}
\end{minipage}
\begin{minipage}{0.49\textwidth}
\centering
\includegraphics[width=\textwidth]{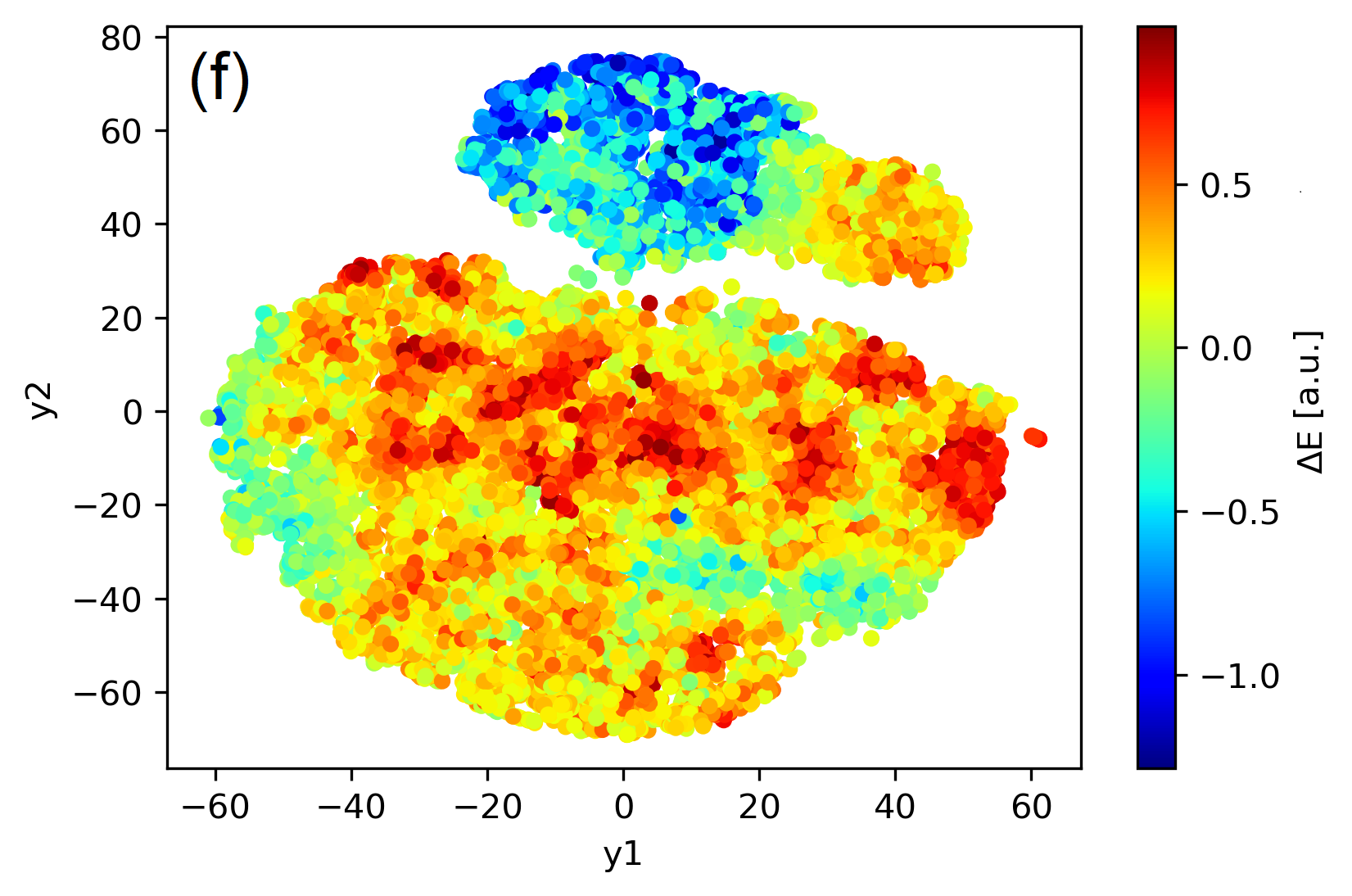}
\end{minipage}\\
\caption{Plot of the 16 995 9-dimensional data points that have been projected onto two axis $\mathrm{y}_1$ and $\mathrm{y}_2$ using the t-SNE dimensional reduction technique described in section \ref{sec:tools}. Each subplot has the projected data plotted as a function an initial input parameter. The initial input parameters include (a) the ionisation rate $\Gamma$, (b) the right molecular well depth $Z_r$, (c) the electron localisation $\alpha$, (d) the field strength $E_0$, (e) the internuclear distance $R$. Also included but not shown here are the left molecular well depth $Z_l$, the left initial wavepacket width $\gamma_l$, the right wavepacket width $\gamma_r$ and the phase difference between the left and right initial wavepackets $\theta$. In subplot (f) the separatrix energy difference $\Delta E$ is shown for each data point, however it is not one of the input parameters. The input data set has been reduced from 100 000 data points to only have either $\Gamma < 0.002$ or $\Gamma > 0.05 $. PCA distributions using the original 100 000 data points are shown in Appendix A.  }
\label{fig:StaticTsne2}
\end{figure*}

As seen in Figure~\ref{fig:StaticTsne2}(a), the algorithm successfully separates two clusters of low ($\Gamma < 0.002$) and high ($\Gamma > 0.05$) ionisation yield. Within the high ionisation cluster, the data points are further distinguished by a very high ($\Gamma > 0.3 $) ionisation yield region. From the distribution of other parameters in  these two clusters, we can determine which lead to a high ionisation yield.

The strongest predictor of high ionisation rate is the nuclear depth factor $Z_r$, [see Fig.~\ref{fig:StaticTsne2}(b)], which in this configuration is associated to the upfield molecular well. As shown in Figure~\ref{fig:rate-contour}, the ionisation rate falls of sharply as soon as $Z_{r} > 1.0$. While the value of $Z_r$ is originally chosen randomly in the range [1:2], the average value of $Z_r$ in the high ionisation yield cluster is 1.16, and 1.12 in the $\Gamma > 0.3 $ region. Indeed, 90\% of  the $\Gamma > 0.3 $ ionisation yield data points have $Z_r < 1.25$. 

In Fig.~\ref{fig:StaticTsne2}(c), echoing the conclusions from Fig.~\ref{fig:tsne1} and Fig.~\ref{fig:relAmpl}, only 4\% of the high ionisation yield data points have the initial wavepacket of the electron localised downfield, with $\alpha > 0.8$. Similarly, 30\% of the  high ionisation yield cluster data points are for an initial wavepacket localised upfield, with $\alpha < 0.2$. This proportion goes up to 72\% for $\Gamma > 0.3 $.

From Fig.~\ref{fig:StaticTsne2}(d), the median initial field strength $E_0$ of the high ionisation yield cluster is 0.07, despite the parameter being originally randomly chosen from a [0.04:0.08] range. This means the majority of high ionisation yield data points have a high field strength. Indeed, 82\% of data points have $E_0 > 0.06~ \mathrm{a.u.}$ in that cluster and 88\% in the $\Gamma > 0.3 $ region.

As hypothesised earlier, the clear internuclear distance peak from Fig.~\ref{fig:tsne1} is now less pronounced in Fig.~\ref{fig:StaticTsne2} (e) as we are now also varying the molecular weights $Z$ and the field amplitude $E_0$. Still, 72\% of data points within the very high, $\Gamma > 0.3 $, ionisation yield region have internuclear distances $R$ in the range $6 < R < 9$. This is surprising as the initial range for $R$ is randomly taken from [3:12]. Therefore, one would expect that one third of the data points would be in this range. 

Finally, in Figure~\ref{fig:StaticTsne2}(f), we can deduce the effect of the separatrices on the ionisation yield. Indeed if the separatrices are nested (see Fig.~\ref{fig:phaseSpace}), meaning that $E_S > E_C$, or $\Delta E <0$, then the ionisation rate is suppressed. In contrast, 99.5\% data points in the high ionisation yield cluster have $\Delta E > -0.5$, and 86\% of them have $\Delta E > 0$. This separation is present despite $\Delta E$ not being used as an input parameter.

In conclusion, the t-SNE enables us to visualise the effect of multiple parameters simultaneously. All the conclusions drawn in section \ref{sec:qualitative} are visualised in Fig.~\ref{fig:StaticTsne2}. Parameters with little to no effect are swiftly identified and separated from relevant parameters. Not only that, but a hierarchy of the effect of different parameters is also established much more rapidly. For example the effect of $Z_l$ on the ionisation rate being greater than that of $\alpha$. Because of this we will be able to use the t-SNE in the following section to rapidly assert what parameters are relevant and to what degree.

\section{Time dependent fields - control of enhanced ionisation - optimising autocorrelation step function}
\label{sec:TD}

When using a polychromatic field (see Eq.~\ref{eq:polychrom}) and specific parameters, we can obtain a `controlled' ionisation release, which translates to a step function in the autocorrelation function plot. 
In the following section we will be looking at what parameters lead to step functions and why specific configurations lead to a controlled ionisation release. 

\subsection{t-SNE application to step functions}
\label{sec:TDtsneResults}
\subsubsection{The 3 clusters and electron localisation.}
\begin{figure*}[tbp]
\centering
\includegraphics[width=\textwidth]{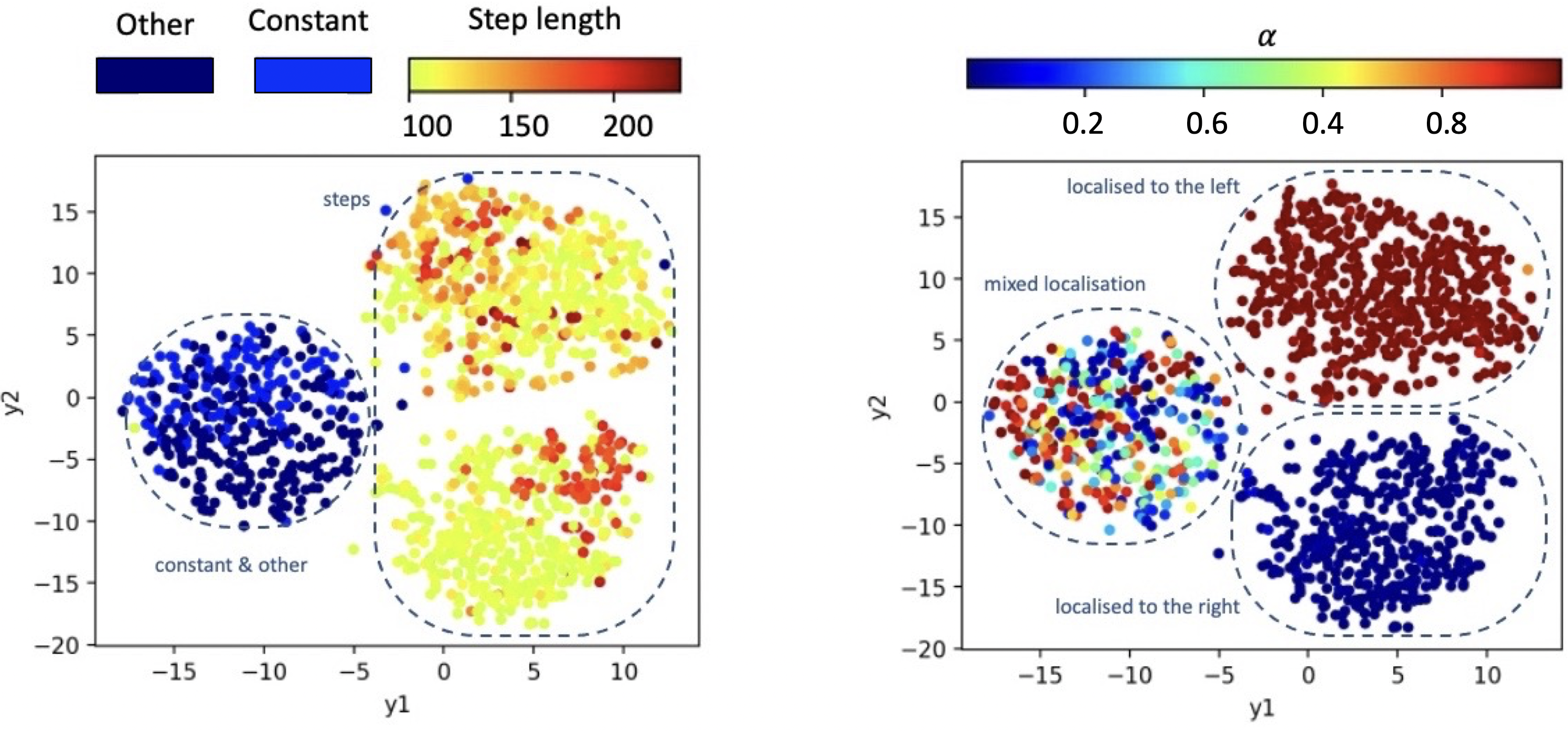}
\caption{Plot of the 1332 12-dimensional data points that have been projected onto two axis $\mathrm{y}_1$ and $\mathrm{y}_2$ using the t-SNE dimensionality reduction technique described in section \ref{sec:tools}. Each subplot has the projected data plotted as a function of an initial input parameter (shown by the colour bar). All units are in [a.u.]. unless stated otherwise. The input parameters shown in this figure are (a)  The step function shape obtained by the sorting algorithm in section \ref{sec:characterisation} that outputs `constant', `other', or the length of the step in the autocorrelation function and (b) The initial electron wavepacket localisation represented by the parameter $\alpha$ in Eq.~(\ref{eq:genwp}). The full list of input parameters as well as details on the initial data set used can be found in section \ref{sec:tools}. PCA distributions using 500 000 data points are shown in Appendix A.}
\label{fig:TDTsneCluster}
\end{figure*}

The t-SNE algorithm separates the data into three different clusters, whose nature can be seen in the two plots of Fig.~\ref{fig:TDTsneCluster}. Therein, the data points are coloured with respect to (a) their autocorrelation function type and (b) their initial electron localisation $\alpha$. Indeed, one cluster groups all autocorrelation functions classified as `constant' or `other', while the other two contain the autocorrelation functions classified as `steps'. Those two clusters are understood by looking at the distribution with respect to the initial electron localisation. All autocorrelation functions that are classified as `steps' have their initial electron wavepacket localised completely at one centre. The clusters are separated between having the initial electron wavepacket localised at the left centre ($\alpha > 0.95$, or dark red) and localised at the right centre ($\alpha < 0.05$, or dark blue). From this it is clear that the initial wavepacket must be localised either to the left or to the right for there to be steps. This is explained when looking at Fig.~\ref{fig:localisation2}, in section \ref{sec:TDeLoc}.
\subsubsection{Difference between the two `step' clusters and the `other' and `constant' cluster. }
\begin{figure*}[tbp]
\centering
\includegraphics[width=\textwidth]{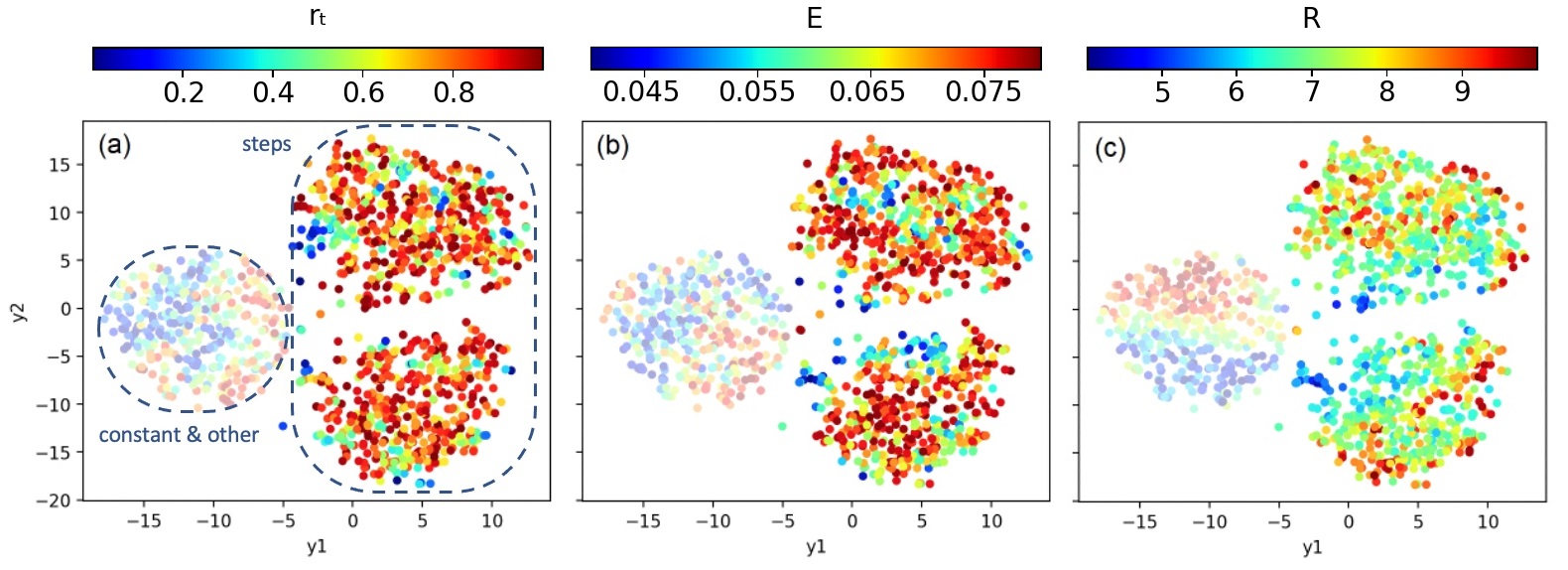}
\caption{Plot of the 1332 12-dimensional data points that have been projected onto two axis $\mathrm{y}_1$ and $\mathrm{y}_2$ using the t-SNE dimensional reduction technique described in section \ref{sec:tools}. Each subplot has the projected data plotted as a function an initial input parameter (shown by the colour bar). All units are in [a.u.]. unless stated otherwise. The input parameter shown in this figure are (a) The ratio $r_t$ between the two colour field amplitudes, (b) the overall field amplitude $E_0$ in Eq.~(\ref{eq:polychrom}) and (c) the internuclear distance $R$. The full list of input parameters as well as details on the initial data points used can be found in section \ref{sec:tools}. PCA distributions using 500 000 data points have been checked for accuracy (not shown).}
\label{fig:TDTsneStepParam}
\end{figure*}
From looking at the different clusters, we can find other correlations, shown in Fig.~\ref{fig:TDTsneStepParam}, between certain parameters and the presence of step functions. most notably, the field ratio $r_t$ between the two driving waves, in subplot (a). Indeed, while originally the value of $r_t$ is chosen randomly from a range [0:1], the two step clusters have a median $r_t$ value of 0.81 and an average of 0.75. Moreover, longer step lengths (more than 150 a.u., or orange/red) have a $r_t$ median value of 0.85 and an average of 0.8.

Next, the overall electric field strength $E_0$, chosen at random from a range of [0.04:0.08], has an median and average value of $E_0 = 0.07~\mathrm{a.u.}$. These can be understood as by seeing that both a high field ratio $r_t$ and a high field strength $E_0$ lead to a higher field peak during which more of the population will escape, causing a deeper drop in the autocorrelation function.

Finally, the internuclear distance is clearly closer to the peak internuclear distance of $R = 7~\mathrm{a.u.}$, which is also confirmed by the standard deviation being only 1.05 instead of 1.3. $R$ is also for $92.6 \%$ of all data in the step clusters above a minimum internuclear distance of $R = 6~\mathrm{a.u.}$. The cause of this effect is more complex, and is discussed in detail in section \ref{sec:TDsep}.

\subsubsection{Symmetry between the `localised to the left' and `localised to the right' step clusters. }
\begin{figure*}[tbp]
\centering
\includegraphics[width=\textwidth]{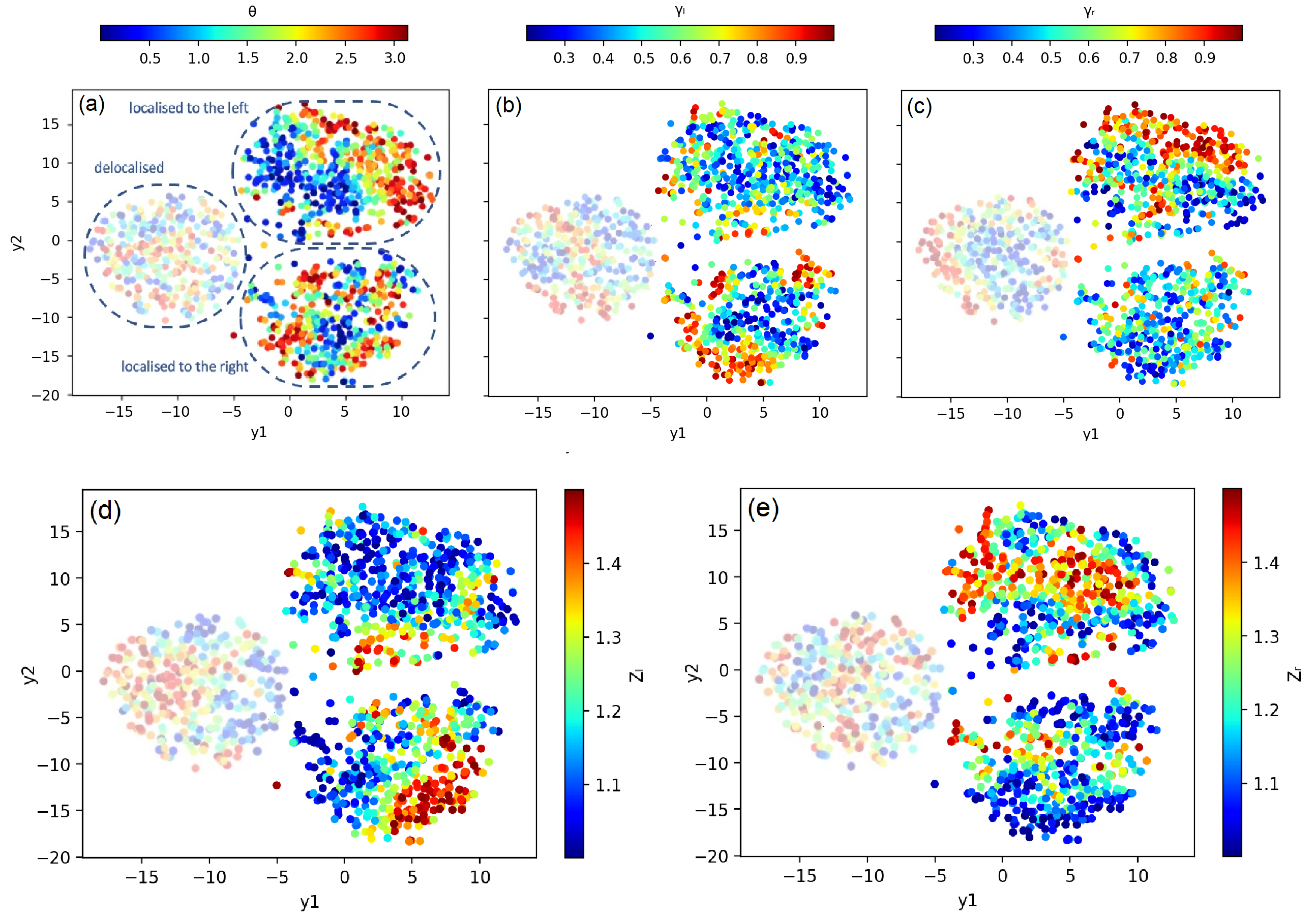}
\caption{Plot of the 1332 12-dimensional data points that have been projected onto two axis $\mathrm{y}_1$ and $\mathrm{y}_2$ using the t-SNE dimensionality reduction technique described in section \ref{sec:tools}. Each subplot has the projected data plotted as a function an initial input parameter (shown by the colour bar). All units are in [a.u.]. unless stated otherwise. The input parameters shown in this figure are (a) the phase difference between the left and right initial wavepackets $\theta$, (b) the left initial wavepacket width $\gamma_l$, (c) the right wavepacket width $\gamma_r$, (d) the left molecular well depth $Z_l$ and (e) the right molecular well depth $Z_r$. The full list of input parameters as well as details on the initial data points used can be found in section \ref{sec:tools}. PCA distributions using 500 000 data points have been checked for accuracy (not shown).}
\label{fig:TDTsneLeftRightSym}
\end{figure*}
The value of $\alpha$ is not the only difference between the two step clusters. The next series of parameters, shown in Fig.~\ref{fig:TDTsneLeftRightSym}, are connected to the symmetry between the left and right molecular wells. First, there is quite naturally the distribution of $\gamma_l$ and $\gamma_r$. From their definition in Eq.~(\ref{eq:gaussian_i}), $\gamma_l$ only influences the initial wavepacket localised to the left. Therefore, if the initial wavepacket was localised to the left, only the initial left wavepacket width $\gamma_l$ is a relevant parameter, and vice versa. Since, as shown in Fig.~\ref{fig:TDTsneCluster}, both step clusters use initially localised wavepackets exclusively, only one $\gamma$ parameter influences each cluster and the $\theta$ parameter (subplot (a)) has no effect on them. As seen in Fig.~\ref{fig:TDTsneLeftRightSym} (b) and (c), 22.8\% of $\gamma_l$ in the localised right step cluster are above 0.8, but only 6.8\%  are above 0.8 in the localised left step cluster. Similarly, 27.5\% of $\gamma_r$ in the localised left step cluster are above 0.8, while 5.3\% are above 0.8 in the localised right step cluster. The width $\gamma = 0.5$ corresponds to the minimal ground-state energy of a field free single-centre soft-core potential with our current parameters, and the vast majority of step functions have a $\gamma$ value that is within the range [0.2 : 0.8]. As $\gamma$ deviates from 0.5, part of the initial wavepacket bleeds out of the bound region despite being initially localised. This leads to oscillations during the flat `step' portion of the autocorrelation function and explains the absence of initial wavepacket widths $\gamma$ above 0.8 in the step clusters. 

The other set of parameters symmetric with respect to the left/right direction are the molecular weights $Z_l$ and $Z_r$. Their effect on one cluster should be mirrored on the other. From Figs.~\ref{fig:TDTsneLeftRightSym}(d) and (e), while there is no clear connection between the localised left cluster and $Z_r$, $Z_l$ is more likely to be close to 1.0 (and vice-versa). Indeed there are 42\% of data points that have $Z_l < 1.2$ in the localised right cluster, but 70\% in the localised left cluster. Similarly, 39\% of data points have $Z_r < 1.2$ in the localised left cluster, but 67\% in the localised right cluster. This is further investigated in section \ref{sec:TDsep}.

\subsubsection{Link between different parameters and the step length.}
\begin{figure*}[tbp]
\centering
\includegraphics[width=\textwidth]{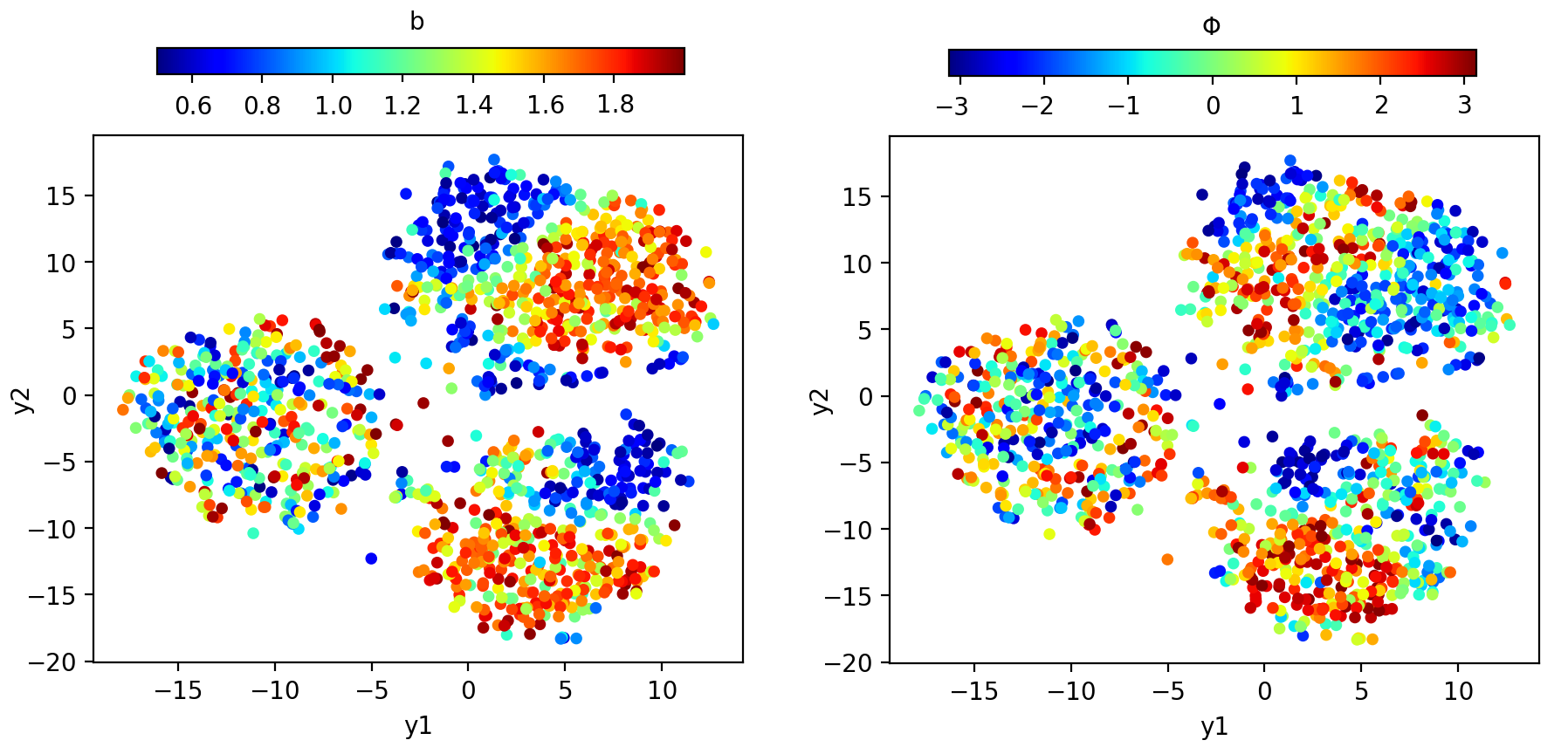}
\caption{Plot of the 1332 12-dimensional data points that have been projected onto two axis $\mathrm{y}_1$ and $\mathrm{y}_2$ using the t-SNE dimensionality reduction technique described in section \ref{sec:tools}. Each subplot has the projected data plotted as a function of an initial input parameter (shown by the colour bar). All units are in [a.u.] unless stated otherwise. The input parameters shown in this figure are (a) The field-frequency ratio $b$ and (b) the field offset $\phi$ in Eq.~(\ref{eq:polychrom}). The full list of input parameters as well as details on the initial data points used can be found in section \ref{sec:tools}. PCA distributions using 500 000 data points have been checked for accuracy (not shown).}
\label{fig:TDTsneStepLength}
\end{figure*}

Going back to the distribution with respect to the autocorrelation function type in Fig.~\ref{fig:TDTsneCluster}, those two `step' clusters are then also distinguished into regions of `short' (less than 150 a.u., or lime yellow) and `long' (more than 150 a.u., or orange/red) step length. 
From Fig.~\ref{fig:TDTsneStepLength},  we can already glean that the field-frequency ratio $b$ affects the step length. However, there is no clear link with the field offset $\phi$. This is further expanded upon in section \ref{sec:TDfreq}. 

\subsection{Analysis of the step functions}
\label{sec:TDfullAnalysis}

In order to obtain a controlled burst of ionisation, two pre-requisites are needed: First, a time interval in which the population stays bounded, and, second, a short burst of enhanced ionisation. Following the conclusions drawn from section \ref{sec:TDtsneResults}, we already know what parameter range is required or preferred. In this section, we will provide an intuitive, physics-based analysis for the controlled-ionisation burst conditions.

\subsubsection{Electron localisation.}
\label{sec:TDeLoc}

\begin{figure}
    \centering
    \includegraphics[width=\linewidth]{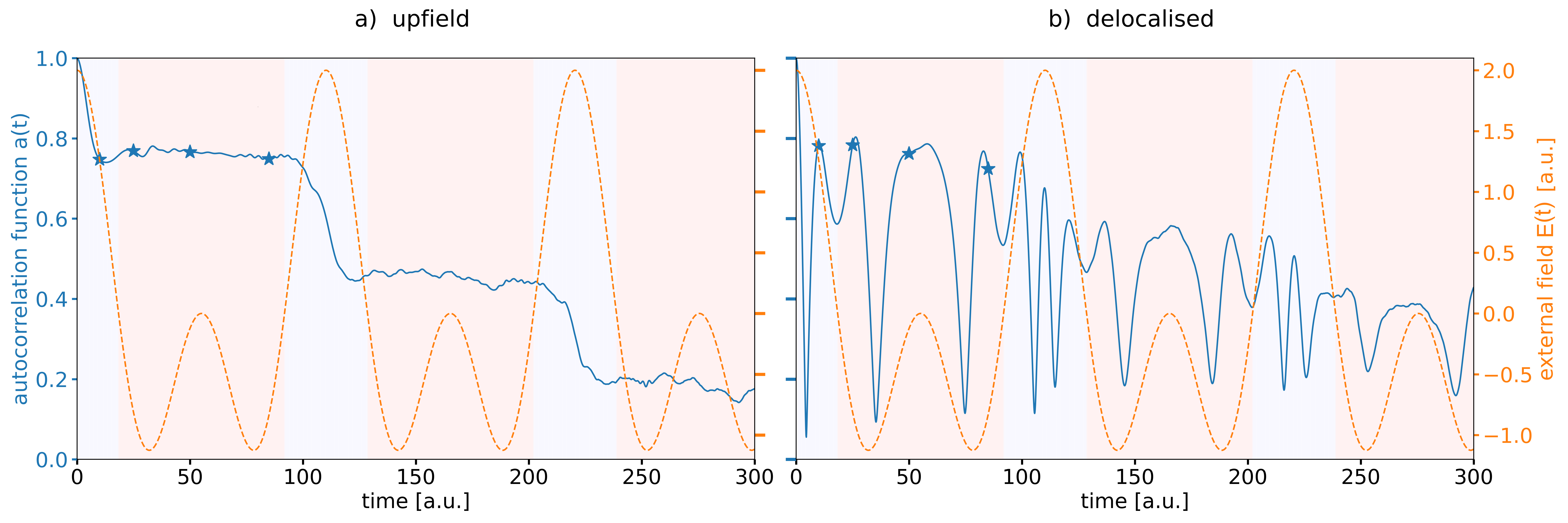}
    \caption{Comparison of auto-correlation functions at R = 6.8 a.u. with initial wavepacket localised upfield or delocalised. Direction of the field is denoted by the colour of the background, blue regions are positive values, pink are negative. The field is shown as the orange dashed lines with values shown on the rightmost y-axis and has form $E \propto \cos{\omega t} + \cos{2 \omega t}$ where $ \omega = 0.057$ a.u. The stars denote autocorrelation function values at times $t = 10~ \mathrm{a.u.}$, $t = 25~ \mathrm{a.u.}$, $t = 50~ \mathrm{a.u.}$, and $t = 85~ \mathrm{a.u.}$.}
    \label{fig:localisation1}
\end{figure}

\begin{figure}
    \begin{minipage}{0.80\textwidth}
    \centering
    \includegraphics[width=\linewidth]{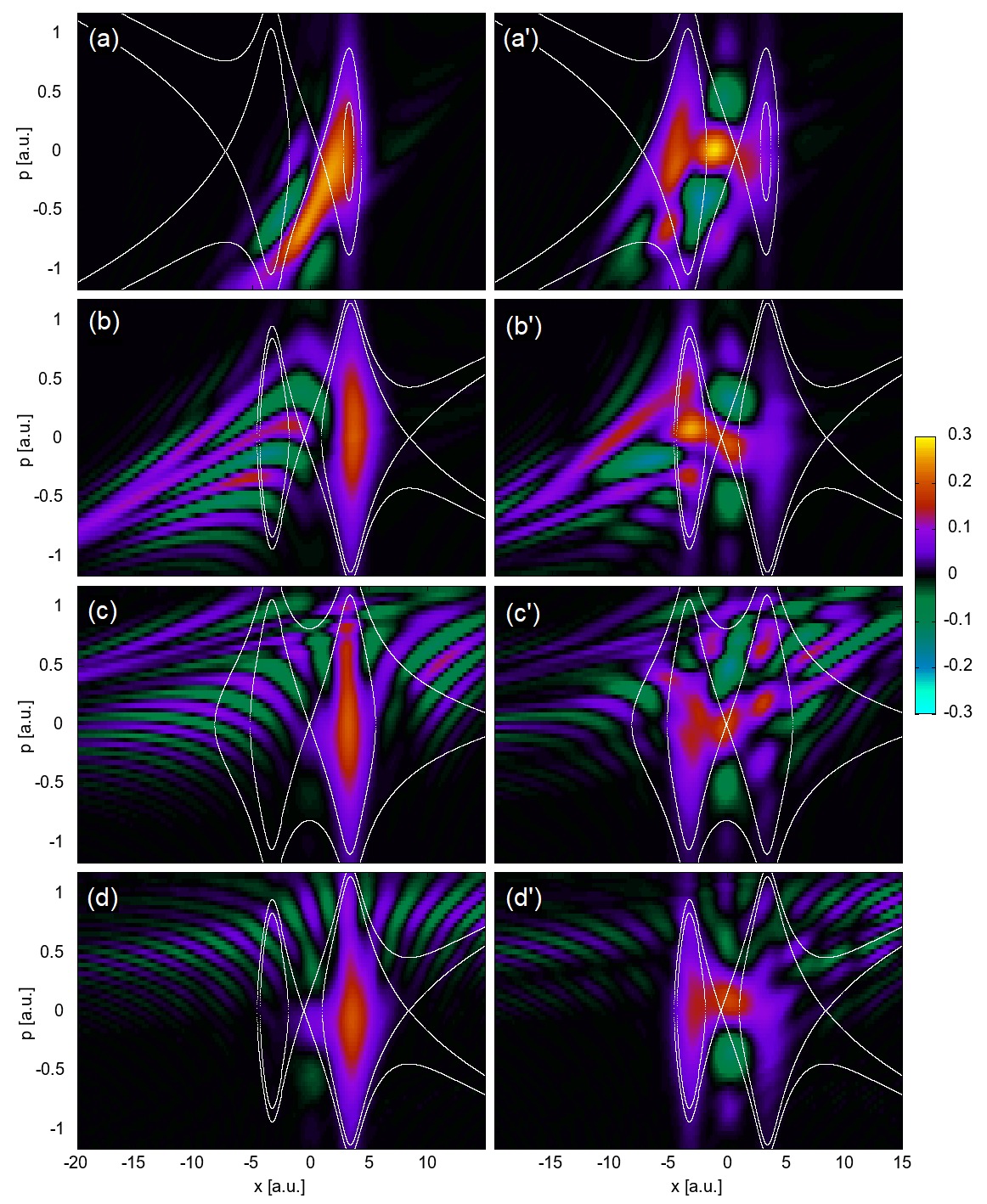}
    \end{minipage}\\
    \caption{ Wigner quasiprobability distributions computed for internuclear distances  R = 6.8 a.u. with initial wavepacket localised at the right potential well ($\alpha = 0.0$)  [left column] or delocalised between the two molecular wells ($\alpha = 0.5$)[right column]. The external field has frequency $ \omega = 0.057$ a.u. and strength $E_0 = 0.0534 ~\mathrm{a.u.}$. The snap shots are taken at times $t = 10~ \mathrm{a.u.}$ for (a) and (a'), $t = 25~ \mathrm{a.u.}$ for (b) and (b'), $t = 50~ \mathrm{a.u.}$ for (c) and (c'), and $t = 85~ \mathrm{a.u.}$ for (d) and (d'). These times are indicated by stars in Fig.~\ref{fig:localisation1}.}
    \label{fig:localisation2}
\end{figure}

From Fig.~\ref{fig:TDTsneCluster}, it is clear that the initial electron wavepacket must be localised, either around the left or right molecular well. Fig.~\ref{fig:localisation1} illustrates the difference between using an initial localised wavepacket and a delocalised wavepacket. This, along with the Wigner function shown in Fig.~\ref{fig:localisation2}, allows us to understand the mechanism behind the controlled ionisation release.

To obtain a burst of enhanced ionisation, the system must be in an optimal configuration. This was initially found in a study using static fields in \cite{chomet2019quantum}, and further expanded upon in section \ref{sec:static}. A key point is that the initial wavepacket must be localised upfield. This allows the population to escape directly through the quantum bridge to the semiclassical escape pathway and ultimately the continuum. It also stops the quantum bridge from cycling through the momentum space and bringing the population back to the upfield centre. Similarly, the population must stay bounded for an interval of time, meaning that the system must be in the configuration with lowest ionisation rate. As seen in Fig.~\ref{fig:relAmpl}, for static fields, this happens when the initial wavepacket is localised downfield.  For the time dependent polychromatic field, the terms `upfield' and `downfield' become relative as the configuration changes with the direction of the field. However in either case the initial wavepacket must be localised on one of the molecular centres. As seen in the behaviour of the Wigner function in Fig.~\ref{fig:localisation2} [right column] these two conditions are not possible if the initial wavepacket is delocalised. Because of the quantum interference in between the two molecular centres, there is always the cyclic motion of the quantum bridge.


The localisation of the wavepacket must work in conjunction with the sign of the external field for the stepwise behaviour to be present in the autocorrelation function. If the wavepacket is localised to the left and the external field $E(t)$ is positive, or if the wavepacket is localised to the right and $E(t)$ is negative, this will result in downfield localisation. Alternatively, upfield localisation occurs for a wavepacket localised to the left and negative $E(t)$, or a wavepacket localised to the right and positive $E(t)$. 

In order to obtain a step function, the localisation must be upfield during the drop and downfield during the flat part of the step. In Fig.~\ref{fig:localisation1}, the different field signs are shown by the different coloured shaded areas. Indeed, during the initial drop, the external field is positive and the population localised to the right, and therefore upfield. It escapes rapidly through the quantum bridge as seen by the Wigner function in Fig.~\ref{fig:localisation2}(a) and the separatrices are open. As soon as the field changes sign, the approximately constant behaviour in the autocorrelation function begins.  As seen in Figs.~\ref{fig:localisation2}(b), (c) and (d), during that time the wavepacket localisation is now downfield. Because of that, despite the separatrices being sometimes open, like in subplots (b) and (d), ionisation is suppressed. When the field is no longer strongly negative around $t = 60$, the separatrices are closed, as shown in subplot (c). This also stops the population from escaping. The autocorrelation function therefore has a step until the field becomes positive again, with the separatrices strongly open. From this analysis and the clustering obtained by the t-SNE distribution, it is clear the electron localisation is the most important parameter in obtaining a controlled ionisation release. 

\subsubsection{Separatrices.}
\label{sec:TDsep}

\begin{figure*}[tbp]
\centering
\includegraphics[width=\textwidth]{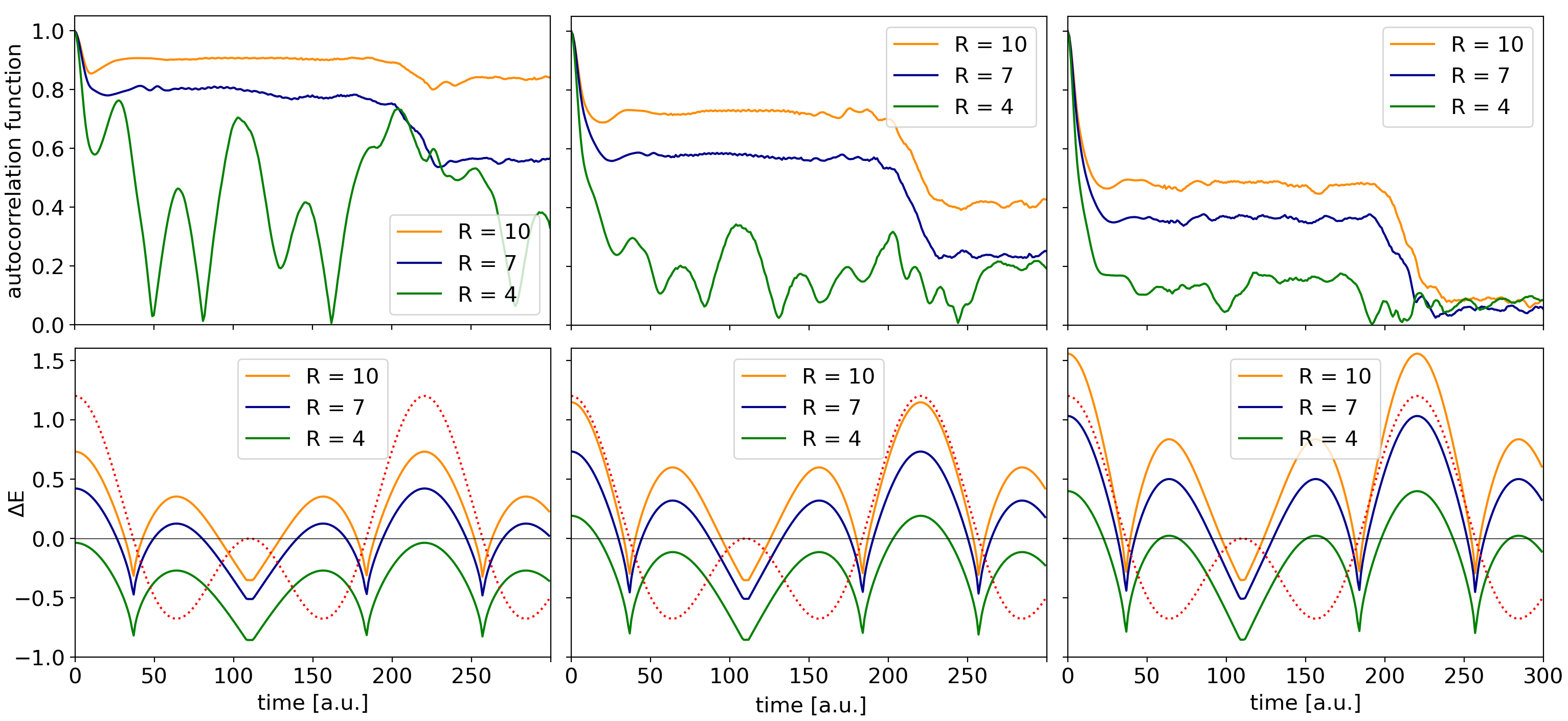}
\caption{Different plots of autocorrelation functions [Top row] and $\Delta E$ [Bottom row] for $R = 4~\mathrm{a.u.}$, $R = 7~\mathrm{a.u.}$ and $R = 10~\mathrm{a.u.}$ with [Left column] $E_0 = 0.04~\mathrm{a.u.}$, [Middle column] $E_0 = 0.06~\mathrm{a.u.}$ and [Right column] $E_0 = 0.08~\mathrm{a.u.}$. The external laser field is represented by the red dotted line (not to scale). We consider an initial wavepacket localised to the right, with $\gamma_r$ = 0.5, and equal 
molecular weights $Z_r = Z_l = 1.0$. The external field amplitude and frequency ratios are $r_t = 1.0$ and $b = 0.5$, respectively. }
\label{fig:TdSepER}
\end{figure*}

\begin{figure*}[tbp]
\centering
\includegraphics[width=\textwidth]{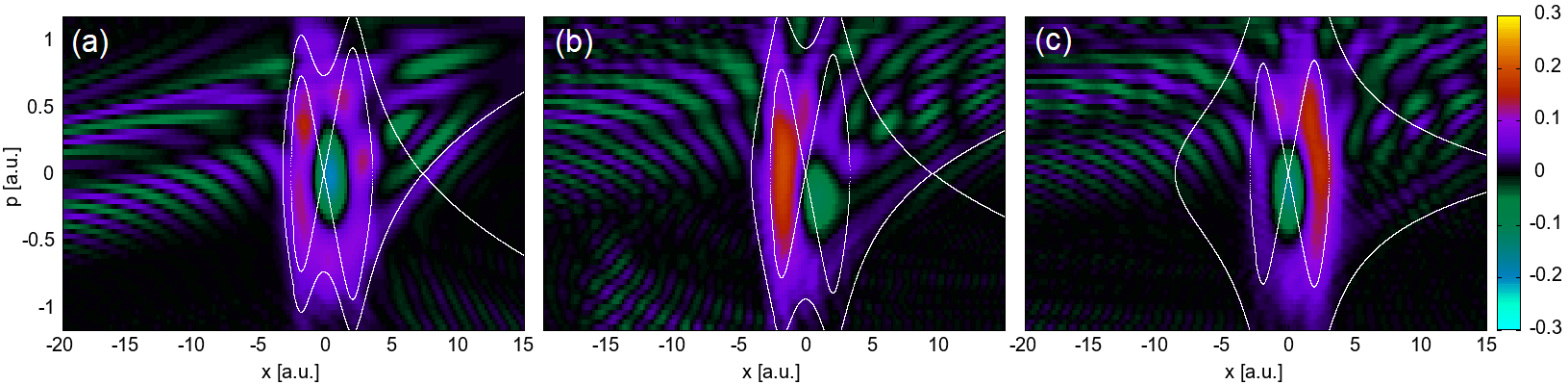}
\caption{Evolution of the Wigner function with internuclear distance $R = 4~\mathrm{a.u.}$ and external field parameters $r_t = 1.0$, $b = 0.5$ and $E_0 = 0.04~\mathrm{a.u.}$. The snap shots are taken at times (a) $t = 65~ \mathrm{a.u.}$, (b) $t = 85~ \mathrm{a.u.}$ and (c) $t = 100~ \mathrm{a.u.}$. Initial wavepacket localised to the right, with $\gamma_r$ = 0.5. Molecular weights $Z_r = Z_l = 1.0$.}
\label{fig:TdSepWig4}
\end{figure*}

\begin{figure*}[tbp]
\centering
\includegraphics[width=0.7\textwidth]{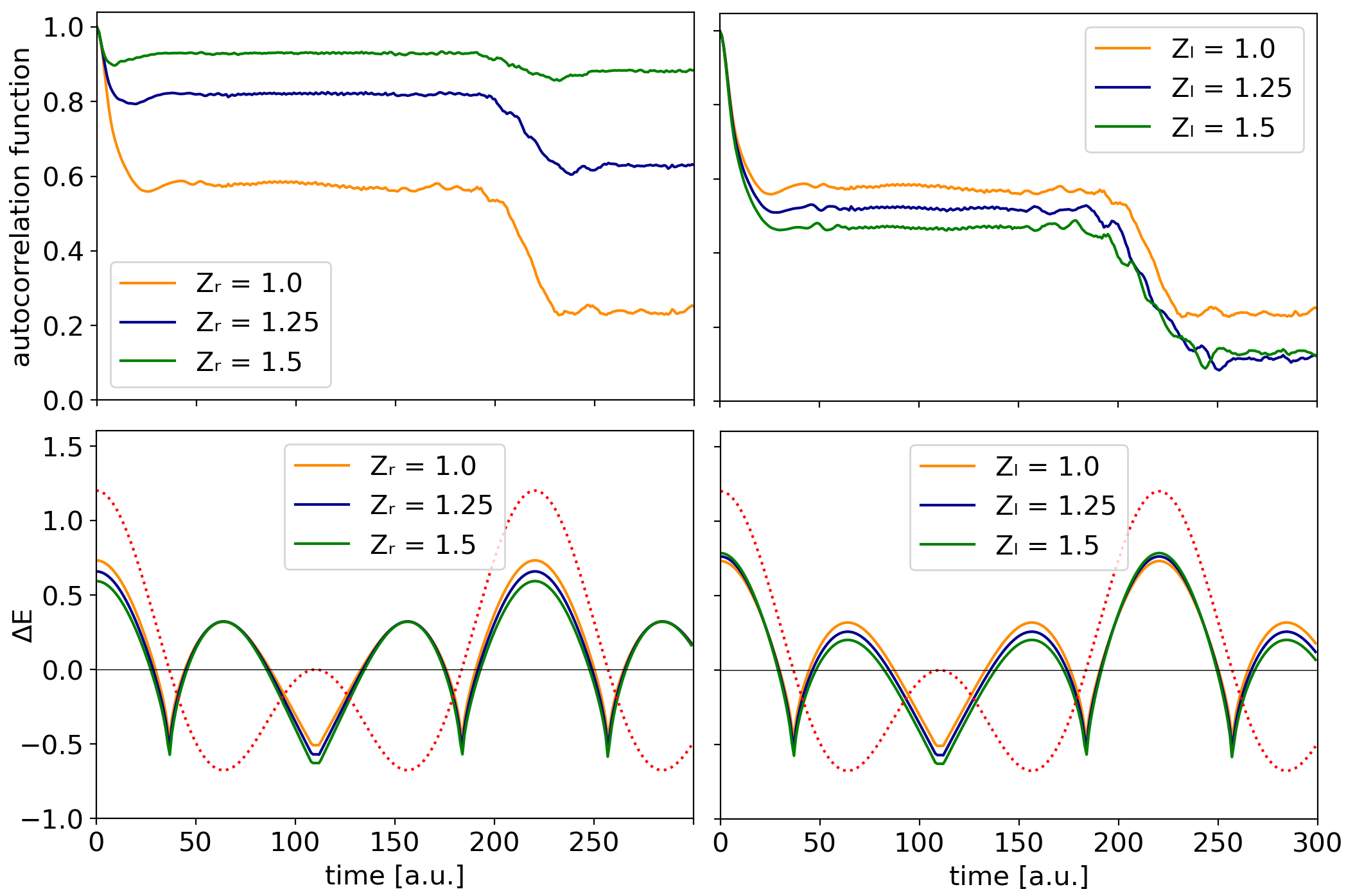}
\caption{Different plots of autocorrelation functions and $\Delta E$ for [Left column] $Z_r = 1.0~\mathrm{a.u.}$,  $Z_r = 1.25~\mathrm{a.u.}$ and  $Z_r = 1.5~\mathrm{a.u.}$ with $Z_l = 1.0~\mathrm{a.u.}$. and [Right column] $Z_l = 1.0~\mathrm{a.u.}$,  $Z_l = 1.25~\mathrm{a.u.}$ and  $Z_l = 1.5~\mathrm{a.u.}$ with $Z_r = 1.0~\mathrm{a.u.}$. The initial wavepacket is localised to the right and other parameters are set to $R = 7~\mathrm{a.u.}$, $E = 0.06~\mathrm{a.u.}$, $r_t = 1.0$ and $b = 0.5$. The external laser field is represented by the red dotted line (not to scale).}
\label{fig:TdSepZ}
\end{figure*}

On top of the electron localisation, the presence of step functions is strongly dependent on the value of $E_0$, $r_t$, $Z_l$, $Z_r$ and $R$, as seen in Fig.~\ref{fig:TDTsneStepParam} and Fig.~\ref{fig:TDTsneLeftRightSym}. Both $E_0$ and $r_t$ have a similar effect on the external laser field: when $r_t$ is close to 1.0 and when $E_0$ is large, the field peak increases in absolute value. As shown in Fig.~\ref{fig:StaticTsne2}, when using a static field, increasing $E_0$ leads to higher ionisation yield. In our time-dependent case, when $E_0$ or $r_t$ is high, the value of $\Delta E$ at the field peak is much higher, meaning the separatrices are wide open. This phenomenon is best seen in Fig.~\ref{fig:TdSepER} for a fixed value of $R$. In the second row it is clear that the increase in $E_0$ leads to higher maximum values of $\Delta E$. For example, at internuclear distance $R = 10$, $E_0 = 0.04$ has $\Delta E = 0.7$ at the field maximum, while $E_0 = 0.08$ has $\Delta E = 1.6$. Compared to the corresponding autocorrelation functions in the first row, we see that when matched by the correct electron localisation, this will lead too a much steeper drop during the decreasing part of the autocorrelation function step. Using our previous example, at $E_0 = 0.04$ the first step plateaus at 0.9, while $E_0 = 0.08$ at 0.5.

The effect of the internuclear distance $R$ is at a glance not clear. When focusing on the initial drop, the connection between $\Delta E$ and the initial drop of the correlation seem contradictory. Intuitively one would expect a that a higher maximum $\Delta E$ lead to a steeper drop of the autocorrelation function, as was the case for varying $r_t$ and $E_0$. But the opposite appears true with varying $R$. 
This is understood by results in \cite{chomet2019quantum}, and illustrated in Fig.~\ref{fig:TdSepER}. As $R$ is becomes too large, the quantum bridge between the two molecular wells weakens and the ionisation rate falls. As consequence the initial drop of the autocorrelation function is low. With $E_0$ being low as well, this leads to a `constant' autocorrelation function as seen in Fig.~\ref{fig:TdSepER} for $R = 10$ and $E = 0.04$. 
When $R$ is too small, the quantum bridges between the molecular wells are at their strongest, and initially facilitate population transfer, leading to a steeper drop. However, as seen in section \ref{sec:static}, a small $R$ does not lead to a high ionisation rate. That is understood in results from \cite{chomet2019quantum}, and illustrated by the Wigner functions in Fig.\ref{fig:TdSepWig4}: the quantum bridge brings population back to the right-side core. This cyclic motion does not follow the direction of the external field. As a consequence, the quasiprobability distribution does not stay within one centre during the `flat' portion of the step autocorrelation function, as illustrated in Fig.\ref{fig:TdSepER}.

While the effect of  $Z_l$ and $Z_r$ differs greatly depending on the step cluster analysed, it is mirrored, so conclusions on one can be expanded to both. Focusing on the initially localised to the right cluster from Fig.\ref{fig:TDTsneLeftRightSym}, we see that while $Z_l$ has little influence, the vast majority of step autocorrelation functions has $Z_r < 1.2$. This is illustrated in Fig.\ref{fig:TdSepZ}: The initial drop of the step autocorrelation function greatly reduces as $Z_r$ increases. From the corresponding $\Delta E$ plot, the peak separatrix energy difference goes from 0.7 for $Z_r$ = 1.0 to 0.6 for $Z_r$ = 1.5. Compared to the effect of the separatrix energy difference of $E_0$ and its effect on the initial autocorrelation function drop, it appears something else must be at play. A better understanding is obtained when looking at the ionisation rate in static fields as a function of $Z_r$ and $Z_l$. As explained in section \ref{sec:qualitative}, the biggest difference between Fig.~\ref{fig:WignerZrZl} is not the separatrix energy difference, but the range of the bound region, especially in the momentum space.  Back to the time dependent situation, as $Z_r$ increases the population (here initially localised to the right) stays trapped at the right molecular well, and the initial step drop will be much shorter.

\subsubsection{Step length and two colour field frequencies.}
\label{sec:TDfreq}

From Fig.~\ref{fig:TDTsneStepLength} we can already observe that the step length is connected to the value of the two colour field frequency ratio $b$. Indeed, while step functions of step length less than $110~\mathrm{[a.u.]}$ have on average $b = 1.5$, step function of step length greater than $150~\mathrm{[a.u.]}$ have on average $b = 0.9$. 

\begin{figure*}[tbp]
    \centering
    \includegraphics[width=\textwidth]{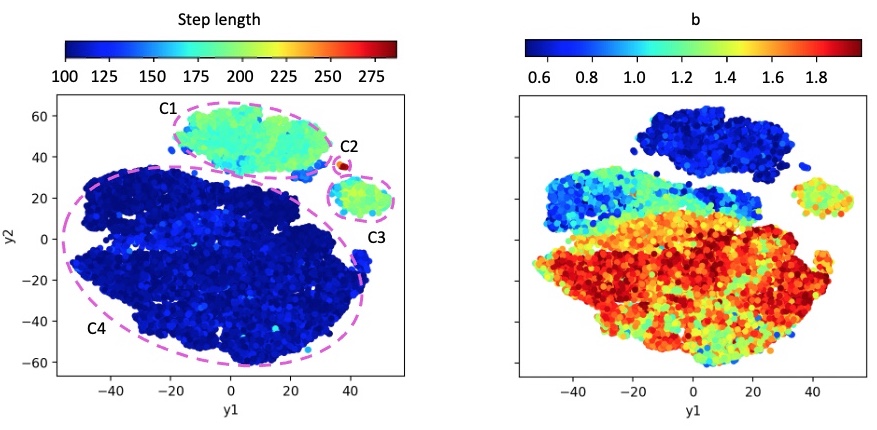}
    \caption{Plot of the 27 194 8-dimensional data points that have been projected onto two axis $\mathrm{y}_1$ and $\mathrm{y}_2$ using the t-SNE dimensional reduction technique described in section~\ref{sec:tools}. All units are in [a.u.]. unless stated otherwise. Each subplot has the projected data plotted as a function a initial input parameter. The initial input parameters include (a) The output of the step function algorithm from section~\ref{sec:characterisation} that is the length of the step in the autocorrelation function and (b) the frequency-ratio of the second colour field $b$ as well as not shown here the offset of the second field $\phi$, the field ratio $r_t$, the internuclear distance $R$, the left molecular well depth $Z_l$, the right molecular well depth $Z_r$, the right initial wavepacket width $\gamma_r$. All data points use an initially localised to the right wavepacket with $E_0 = 0.07$. The input data set has been reduced from 1 000 000 data points to only look at step autocorrelation functions. PCA distributions using the original 1 000 000 data points have been analysed for accuracy (not shown).}
\label{fig:TdFieldTsne}
\end{figure*}

\begin{figure*}[tbp]
    \label{fig:TDfrequency}
    \begin{minipage}{0.49\textwidth}
    \centering
    \includegraphics[width=\textwidth]{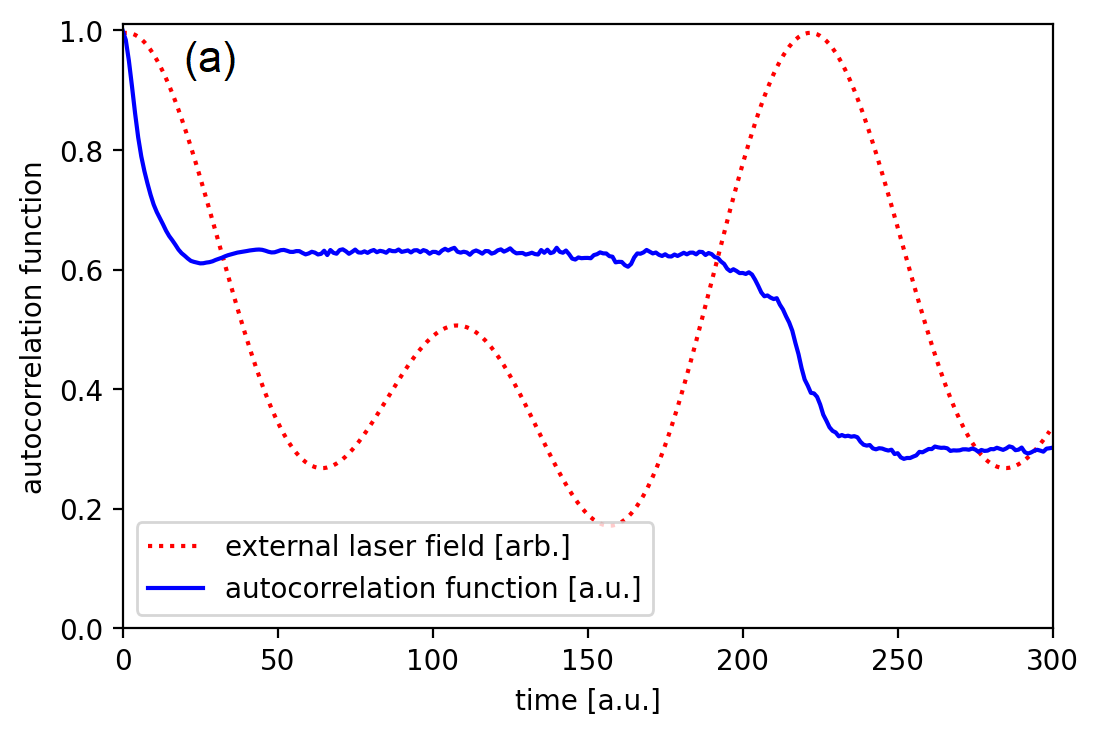}
    \end{minipage}
    \begin{minipage}{0.49\textwidth}
    \centering
    \includegraphics[width=\textwidth]{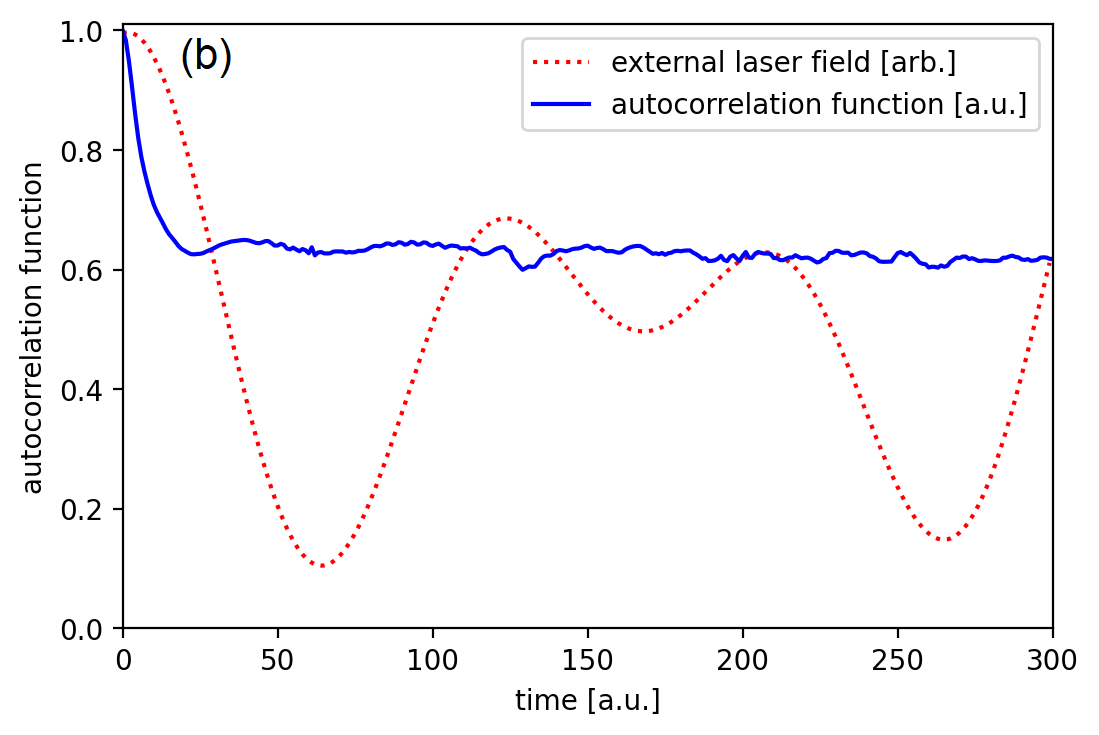}
    \end{minipage}
    \begin{minipage}{0.49\textwidth}
    \centering
    \includegraphics[width=\textwidth]{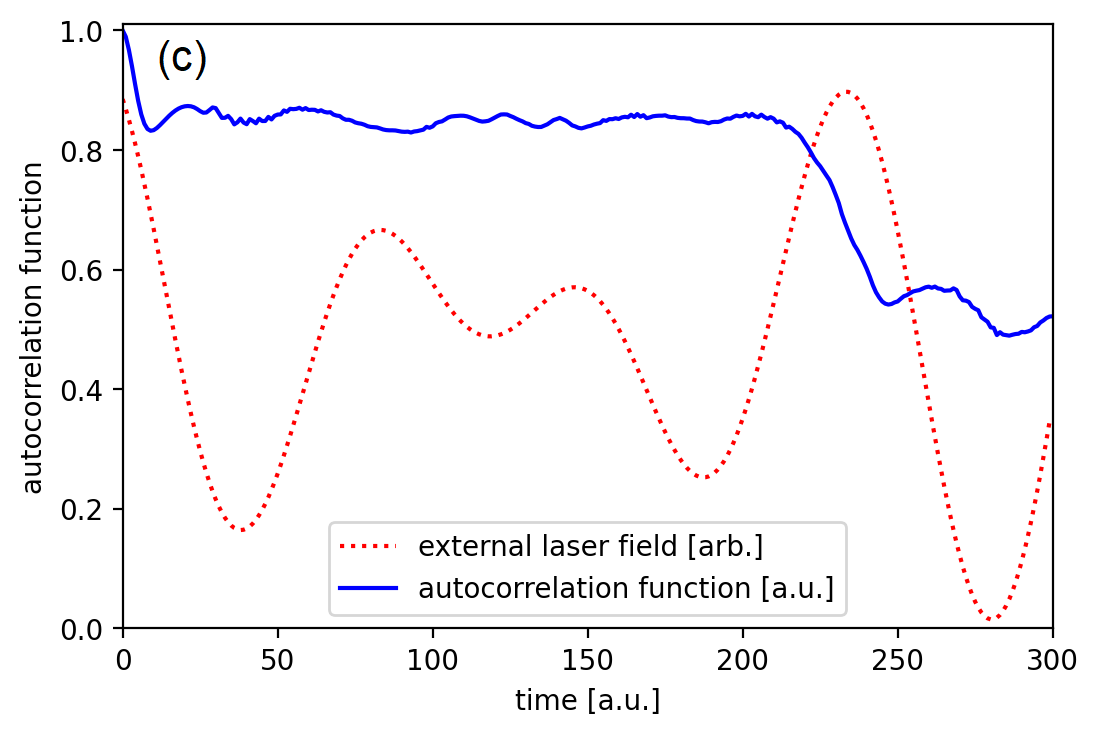}
    \end{minipage}
    \begin{minipage}{0.49\textwidth}
    \centering
    \includegraphics[width=\textwidth]{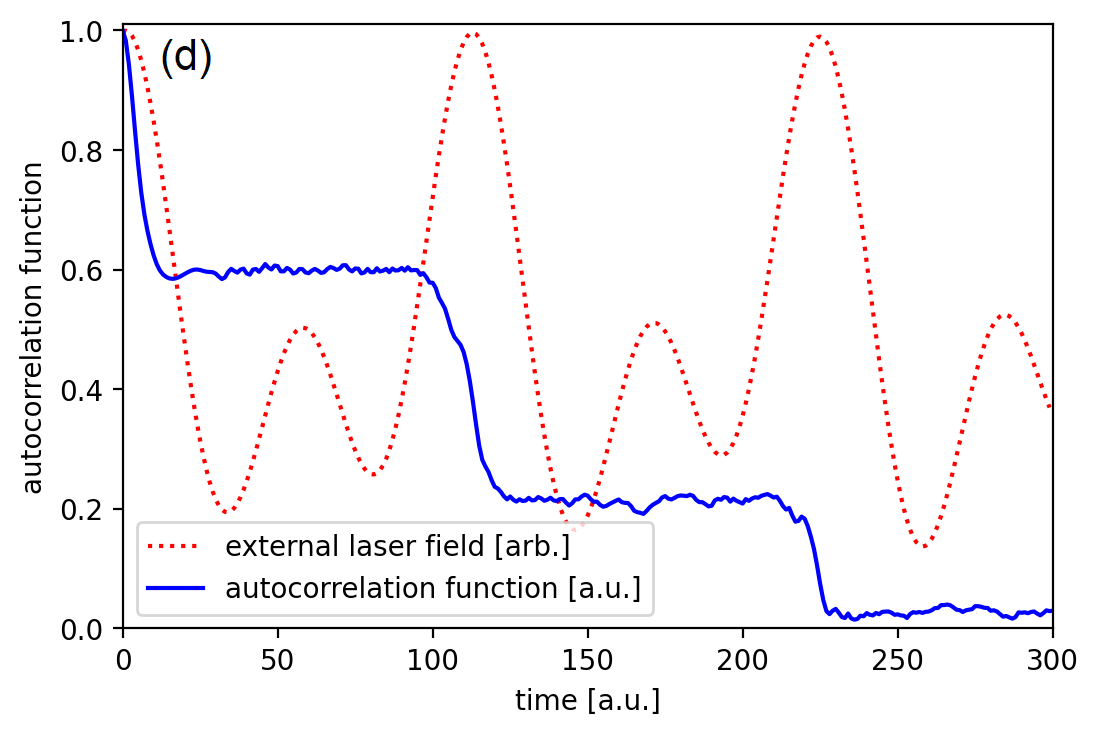}
    \end{minipage}
\caption{Plot of the corresponding autocorrelation function and external field in each of the four different clusters in Fig.~\ref{fig:TdFieldTsne}, in order of increasing field frequency ration $b$. Constant parameters are $E_0 = 0.07$, $R = 7$, $\alpha = 0.0$, $\gamma_r = 0.5$ . The parameters used in each subplot are (a) from C1; $b = 0.5$, $r_t = 1.0$, $\phi = -0.2$, $Z_r = 1.2$ and $Z_l = 1.2$, (b) from C2; $b = 0.7$, $r_t = 1.0$, $\phi = -0.2$, $Z_r = 1.2$ and $Z_l = 1.2$, (c) from C3; $b = 1.3$, $r_t = 1.0$, $\phi = 1.0$, $Z_r = 1.1$ and $Z_l = 1.1$ and (d) from C4; $b = 2.0$, $r_t = 1.0$, $\phi = -0.13$, $Z_r = 1.05$ and $Z_l = 1.5$.}
\label{fig:TdFieldStep}
\end{figure*}
In order to view the relationship between the step length and other parameters in greater detail, we will now fix certain parameters in order to get a clearer picture of the effect the others, as well as only conserve data points with `step' autocorrelation functions. First, we will start with the initial wavepacket localised at the right centre. This is equivalent to focusing only on the `localised to the right' cluster of Fig.~\ref{fig:TDTsneCluster}. Because of this, we also remove the initial wavepacket width $\gamma_l$ and $\theta$ parameters. Finally, we will fix the field strength to $E_0  = 0.07~\mathrm{a.u.}$. These results are shown in Fig.~\ref{fig:TdFieldTsne}.

The t-SNE algorithm separates the data into four different clusters. The C1 and C3 clusters represent step lengths between  $110$ and $260$. They are distinguished by the value of the frequency ratio $b$ of the two-color fields: the largest C1 having an average of $b = 0.65$ and the smallest C3 $b = 1.32$. Next, the C2 cluster is the smallest and corresponds to all autocorrelation functions with a very long (greater than $260$) step length. Finally, the C4 cluster is the largest cluster and corresponds to all autocorrelation functions with a short step length. Here the average step length is $1.09$. An example of autocorrelation function and external field within each cluster is shown in Fig.~\ref{fig:TdFieldStep}. 

As seen in the previous sections, the temporal location of the field peak, when matched with the population localisation, determines the step drop. From this, we can assume that the specific times for which the field peaks occur match the step length. This can be seen in Fig.~\ref{fig:TDfrequency} (a) and (b), when increasing the second field frequency-ratio from $b=0.5$ to $b = 0.7$, the distance between two field peaks increases and the step length goes from 187 to 285. This is also confirmed when looking at an example step function from clusters (c) and (d): The distance between field peaks matches the step length.  

\section{Conclusions}
\label{sec:conclusions}

In this work we use dimensionality reduction techniques, namely the  the t-distributed stochastic neighbor embedding (t-SNE) method and Principal component analysis (PCA) to investigate the effect of multiple parameters at once in enhanced ionisation.  We show how quantum effects in strong-field ionisation of stretched diatomic molecules may be understood, classified and synchronised with the external laser field in order to create tailored ionisation bursts. 
Thereby, the t-SNE was crucial to establish a hierarchy of parameters, and manipulate the relevant time scales.  These time scales are either dictated by the field, or by the molecular system. The latter are associated to quantum-interference structures that provide a direct, intra-molecular population transfer via a quantum pathway in phase space, the quantum bridges \cite{chomet2019quantum,kufel2020alternative}. The results presented here have been mainly obtained with the t-SNE, while the PCA was used to rule out possible algorithm-dependent artefacts.

The effectiveness and accuracy of the t-SNE is first showcased on the ionisation rate using a static field. The expected results are found using phase-space arguments and Wigner quasiprobability distributions. They completely match the conclusions drawn from the t-SNE distributions. Furthermore, a hierarchy in the importance of different parameters is quickly and effectively demonstrated. Because of this, we confidently used the t-SNE to project data of autocorrelation functions and their parameters for time-dependent fields. 
Using a two-colour field, we obtained a controlled ionisation release, which translates to a stepwise behaviour in the autocorrelation function. By using the t-SNE to project multi-dimensional data sets into 2 dimensions, ionisation profiles obtained while varying multiple parameters at once are organised and understood. 

A conclusion from previous work \cite{chomet2019quantum} was that a key to understand enhanced ionisation is the interplay between the intra-molecular quantum bridges and the external field. This also holds for controlled ionisation release (represented by a step function in the autocorrelation function), where the system requires a short burst of enhanced ionisation followed by a time interval for which the population stays bounded.

Quantum bridges are tied to many predictors for a step function in the autocorrelation function, including the strongest one: electron localisation. A delocalised initial wavepacket creates an interference pattern in both positive and negative momentum space. This leads to ionisation bursts that do not follow the external field. Therefore, we cannot obtain a time interval for which the population stays bounded. Moreover, the quantum bridges are highly tied to the internuclear distance. For that reason, configurations that suppress momentum gates do not lead to the burst of ionisation needed. This is the case when the internuclear distance is too large. Alternatively, a too small internuclear distance leads to cyclic motion in the momentum space, regardless of the initial state \cite{chomet2019quantum}. This also stops the population from staying bounded at one centre. 

On the other hand, a higher field peak maximum leads to stronger ionisation bursts. A stronger field peak is achieved by increasing the field amplitude $E_0$ or adding a second field. This is neatly quantifiable by the separatrix energy difference $\Delta E$. From this we determine bounded regions with closed separatrices and ionisation burst regions when the localisation is upfield with open separatrices. The timing of these field peaks matching with the upfield population lead to the start of the autocorrelation function drops. The field frequency determines the distance between the field peaks and is therefore the key to controlling the length of the steps.
Finally, changing the molecular weights $Z_{r,l}$ affects the presence of step functions. As the upfield molecular weight increases, the upfield bound region increases and this greatly suppresses ionisation. 

In summary, we can understand the physical cause and requirements for controlled ionisation release by separating the phenomenon into two steps, a short burst of enhanced ionisation and a time interval where the population stays bounded. For static fields, the optimal configuration for enhanced ionisation is pinpointed to an initial wavepacket localised upfield at optimal internuclear distance with a high intensity external field and a low upfield molecular weight. The separatrices are open with a high separatrix energy difference. This also holds during the time-dependent ionisation burst time interval, and will be periodic as the upfield and downfield configurations will change along with the sign of the external field, matching the field frequency. For the population to stay bounded, either the field strength must be low, separatrices closed or barely open, or the population must be localised in the downfield centre and therefore not follow the changes in the field strength and direction. In addition, the population must not cycle around through momentum gates, meaning the internuclear distance cannot be too short. The t-SNE was greatly effective at separating results into clusters and therefore visualising results due to specific combinations of parameters. For example, an overall analysis of the effect of $Z_r$ on the step function autocorrelation function reveals only negligible effects. However, the step function results were separated into two clusters $\alpha < 0.2$ and $\alpha > 0.8$. The effect of the increased bound region and ionisation suppression from $Z_r$ only affecting the $\alpha < 0.2$ cluster is brought into focus. Moreover, the t-SNE presented a hierarchy of the correlation between different parameters and the quantity of study. From that, the role of electron localisation was pinpointed as the key parameter.

 The importance of electron localisation is in agreement with the findings by several groups, reported over many years. In fact, resolving and ultimately controlling electron localisation in extended or dissociating molecules holds the promise of steering chemical reactions. This has led to many studies, both theoretical and experimental, employing, for instance, pump-probe schemes \cite{Yudin2005,He2007,Sansone2010,Singh2010,Wang2014,Xu2017}, CEP stabilised few-cycle pulses \cite{Roudnev2004,Kling2006}, long-wavelength fields \cite{Liu2011,Liu2013} or synthesised wave forms \cite{Lan2012}. Moreover, orthogonally polarised fields have also been applied to trace or control the ionisation site  \cite{Wu2012,Wu2012b,Liu2017}. Regarding enhanced ionisation, electron localisation in the upfield well, together with coupled charge-resonant states, were widely mentioned in the literature as major contributing factors. This holds both for seminal enhanced ionisation papers \cite{zuo1995charge,Seideman1995} and more recent work in which  models for time-dependent ionisation bursts have been developed \cite{Kelkensberg2011,Takemoto2010,takemoto2011time,Takemoto2011Bohmian}.  

Since potential wells occur in a wide range of physical systems, the present techniques can be applied not only to more realistic molecular models, but also to solids and nanostructures. Thereby, a crucial issue would be to incorporate other degrees of freedom, multielectron dynamics and assess the role of decoherence. Loss of coherence can be caused by many physical mechanisms, such as coupling with additional degrees of freedom, intensity fluctuations, and incoherent emission from across the focal volume (for a brief discussion of some decoherence mechanisms in the context of strong-field quantum sensing see \cite{Maxwell2021}). 
Recent studies using pump-probe schemes in $\mathrm{H}_2$ have shown that coupling of electronic and vibrational degrees of freedom affects coherence and may hinder electron localisation \cite{Vrakking2021}. Still, studies in large molecules indicate that phase relations may be preserved even in systems with many degrees of freedom \cite{Delgado2021}, and could even be controlled leading to tailored ionisation enhancement or suppression \cite{li2020control}. Our simplified model is meant as a proof of concept and a first step towards more realistic scenarios. 

\section*{Acknowledgements} We would like to thank Eduard Oravkin, Pavol Drotar and Samo Hromadka for useful discussions. This work was funded by grant No.\ EP/J019143/1, from the UK Engineering and Physical Sciences Research Council (EPSRC).

\appendix
\section{Principal Component Analysis}
\begin{figure*}[tbp]
\centering
\includegraphics[width=\textwidth]{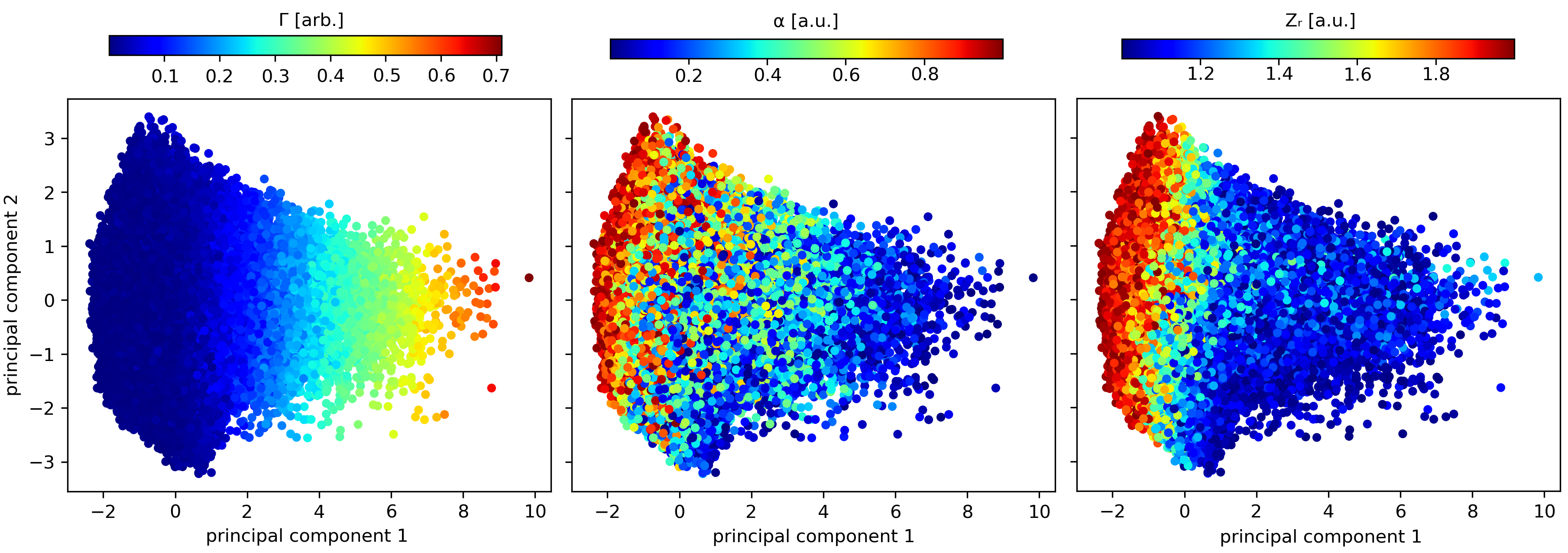}
\caption{Plot of the 100 000 9-dimensional data points that have been projected onto their two principal component axis using PCA. The initial parameters used are equal to that of Figure.~\ref{fig:StaticTsne2}. Each subplot has the projected data plotted as a function an initial input parameter. The initial input parameters include (a) the ionisation rate $\Gamma$, (b) the electron localisation $\alpha$ and (c) the right molecular well depth $Z_r$.}
\label{fig:PCA1}
\end{figure*}

\begin{figure*}[tbp]
\centering
\includegraphics[width=\textwidth]{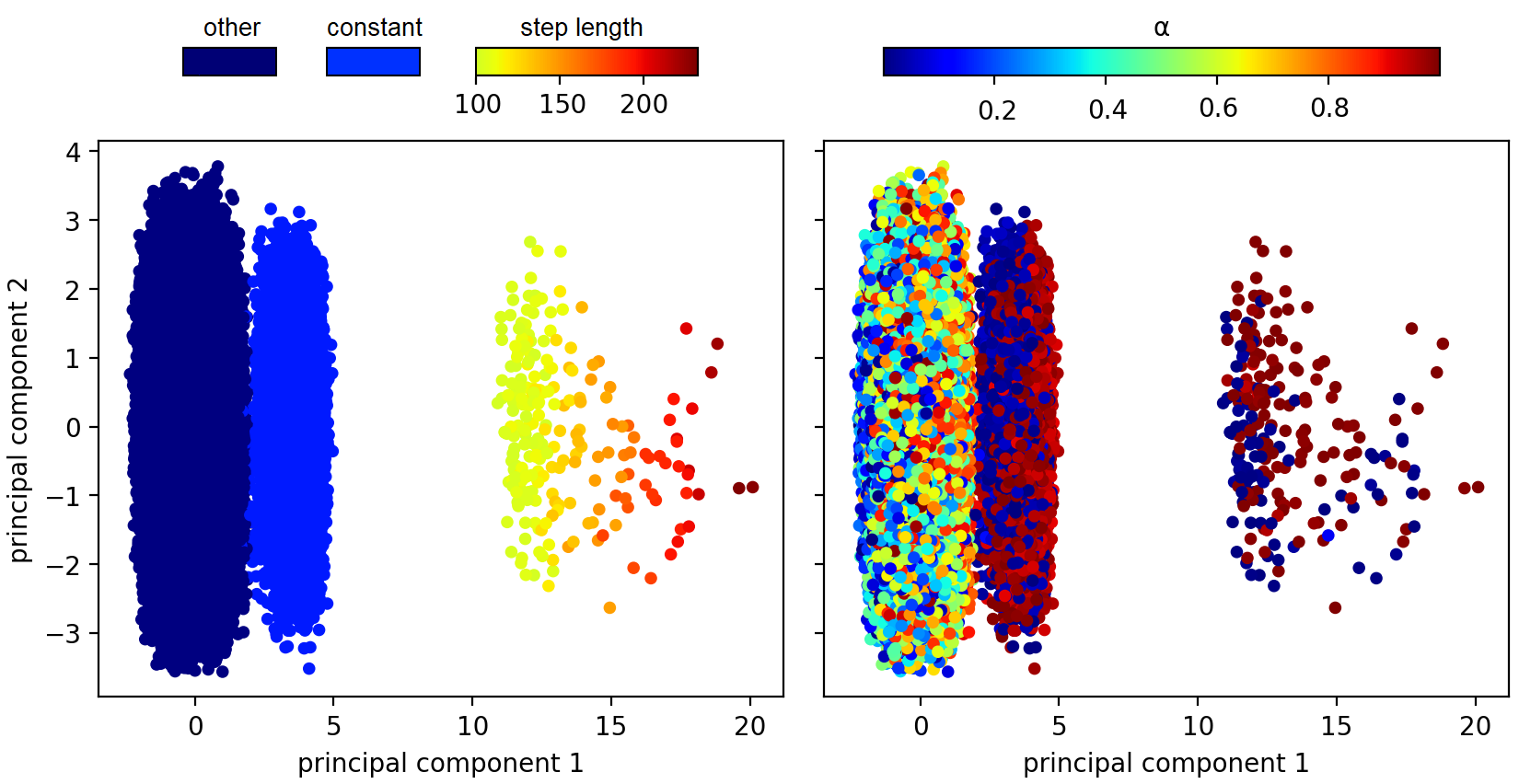}
\caption{  Plot of the 500 000 12-dimensional data points that have been projected onto their two principal component axis using PCA. The initial parameters used are equal to that of Figure.~\ref{fig:TDTsneCluster}. Each subplot has the projected data plotted as a function an initial input parameter. The initial input parameters include (a) The step function shape obtained by the sorting algorithm in section \ref{sec:characterisation} that outputs ‘constant’, ‘other’, or the length of the step in the autocorrelation function and (b) the electron localisation $\alpha$.}
\label{fig:PCA2}
\end{figure*}

In this Appendix we will present a few of the PCA results obtained throughout this study. 
The aim of PCA is to compute the eigenvectors of the covariance matrix (principal components) of the multidimensional data set and recast the data along the two largest eigenvectors.  In other words, the first principal component can be defined as the direction that maximises the variance of the projected data.
Throughout this paper all t-SNE distributions were computed along with PCA distributions for comparison. In this appendix we present two examples, one from the static field analysis in section \ref{sec:static} and one from the time-dependent step autocorrelation function analysis in section \ref{sec:TD}, Fig.~\ref{fig:PCA1} and Fig.~\ref{fig:PCA2} respectively. 

The PCA distribution in Fig.~\ref{fig:PCA1}, obtained for the ionisation rate data set calculated using the static field, does not exhibit individual clusters corresponding to different ionisation rates. However, the distribution is organised following the different ionisation rates. From this we can see that both upfield electron localisation ($\alpha = 0$) and a right molecular weight $Z_r = 1.0$ lead to the highest ionisation rates. 

When it comes to the time-dependent autocorrelation function, the PCA distribution in Fig.~\ref{fig:PCA2} does break into three distinct clusters corresponding to the different autocorrelation outputs: constant, step, and other. This allows us to draw similar conclusions to those obtained with the t-SNE, for instance that the electron must be localised to the right or left in order for the autocorrelation function to have a stepwise shape. However, contrary to the t-SNE distribution shown in Fig.~\ref{fig:TDTsneCluster}, the step autocorrelation initially localised to the right data points and those initially localised to the left are not separated. That means that parameters that affect those two groups differently will not have their effect visually represented, for example $Z_r$ and $Z_l$. 

\printbibliography

%\section*{References}
%\bibliography{iopart-num}
\end{document}